\documentclass[sigconf]{acmart}
\usepackage{subcaption}
\usepackage{multirow}
\usepackage{framed}
\usepackage{graphicx}

\usepackage{acmart-taps}

\usepackage{framed}
\newcommand{\keytakeaway}[1]{
  \begin{framed}
  \noindent\textbf{Takeaway:} #1
  \end{framed}
}

\AtBeginDocument{%
  }

\copyrightyear{2025}
\acmYear{2025}
\setcopyright{cc}
\setcctype{by}
\acmConference[Websci '25]{Proceedings of the 17th ACM Web Science Conference 2025}{May 20--24, 2025}{New Brunswick, NJ, USA}
\acmBooktitle{Proceedings of the 17th ACM Web Science Conference 2025 (Websci '25), May 20--24, 2025, New Brunswick, NJ, USA}
\acmDOI{10.1145/3717867.3717912}
\acmISBN{979-8-4007-1483-2/25/05}

\begin{document}


\title{Public versus Less-Public News Engagement on Facebook: Patterns Across Bias and Reliability}

\author{Alireza Mohammadinodooshan}
\affiliation{%
  \institution{Link\"oping University}
  \country{Sweden}
}
\email{alireza.mohammadinodooshan@liu.se}

\author{Niklas Carlsson}
\affiliation{%
  \institution{Link\"oping University}
  \country{Sweden}
}
\email{niklas.carlsson@liu.se}

\newcommand{\CRedit}[2]{#2}

\begin{abstract}
The rapid growth of social media as a news platform has raised significant concerns about the influence and societal impact of biased and unreliable news on these platforms. While much research has explored user engagement with news on platforms like Facebook, most studies have focused on publicly shared posts. This focus leaves an important question unanswered: how representative is the public sphere of Facebook's entire ecosystem? Specifically, how much of the interactions occur in less-public spaces, and do public engagement patterns for different news classes (e.g., reliable vs. unreliable) generalize to the broader Facebook ecosystem?

This paper presents the first comprehensive comparison of interaction patterns between Facebook's more public sphere (referred to as \textit{public} in paper) and the less public sphere (referred to as \textit{private}). For the analysis, we first collect two complementary datasets: (1) aggregated interaction data for all Facebook posts (public + private) for 19,050 manually labeled news articles (225.3M user interactions), and (2) a subset containing only interactions with public posts (70.4M interactions). Then, through discussions and iterative feedback from the CrowdTangle team, we develop a robust method for fair comparison between these datasets.

Our analysis reveals that only 31\% of news interactions occur in the public sphere, with significant variations across news classes. Engagement patterns in less-public spaces often differ, with users, for example, engaging more deeply in private contexts. These findings highlight the need to examine both public and less-public engagement to fully understand news dissemination on Facebook. The observed differences hold important implications on content moderation, platform governance, and policymaking, contributing to healthier online discourse.

\end{abstract}

\ccsdesc[500]{Information systems~World Wide Web}


\maketitle

\section{Introduction}\label{sec:introduction}

Social media has emerged as an increasingly important news source, with Facebook maintaining a prominent position among social platforms~
\cite{reutersReport2023,pewResearchReport2021}. 
For example, 
according to a 2024 Pew Research Center report~\cite{pewResearchReport2021}, 33\% of U.S. adults regularly get news on Facebook, similar to YouTube (32\%), and notably ahead of Instagram
(20\%) and Twitter (12\%).
This extensive reach highlights the potential impact of news sharing
on Facebook in shaping both individual opinions and broader societal
discourse~\cite{deteminanatsofnewssharing}.

\vspace{8pt}
Unfortunately, not all news are reliable.  With an increasing amount of misinformation being circulated on Facebook, it is crucial to understand how users engage with news of varying reliability~\cite{fake-news-on-facebook}. 
Understanding of
these dynamics are expected to benefit media researchers, journalists, content moderators, and policymakers, whose choices based on these insights, in turn, 
are expected to influence public opinion and the behavior of regular users~\cite{waisbord2018institutional,jones2017social}.

\vspace{8pt}
Previous research has extensively examined public user engagement dynamics with different news content and the factors affecting this engagement ~\cite{edelson2021understanding,aldous2019view,brena2019news,Marwick2017,vitak2011s}. 
However, with a growing privacy inclination of Facebook users and a rise in user engagement in private spaces of Facebook~\cite{facebookusersmoreprivate,lottridge2018let}, it is becoming increasingly important to understand these dynamics also with regards to private sharing. 
Yet, the current literature 
is centered around more public posts, leaving gaps in our understanding of engagement differences between these realms. Most importantly, prior research has not studied the 
dynamics of
public vs. private 
interaction.

\vspace{8pt}
To address this gap,
in this paper
we present the first comprehensive comparison of the news article sharing and user interaction patterns seen on Facebook's more public sphere (referred to as \textit{public} in paper) versus in the less public sphere (for simplicity referred to as \textit{private}).
Specifically, 
we investigate 
differences in
how news articles 
written with different 
{\em bias} and 
{\em reliability} are shared and engaged with, as well as the depth of these interactions.
In this context, {\em bias} refers to a tendency for news articles to exhibit partiality or favoritism towards particular groups or ideas (e.g., left or right on the political spectrum), while {\em reliability} refers to the accuracy and credibility of the information 
\CRedit{presented~\cite{mellado2019perceived}}{presented} 
(e.g., fake or true). 
While most previous research 
on news engagement has primarily focused on either bias or reliability (and on public engagement),
we consider both dimensions, as they have been found related, but yet each has its own ability to influence news article sharing behaviors and affect the quality and diversity of information that users encounter on social 
media~\cite{guess2018less, pennycook2018prior}.

\vspace{8pt}
Furthermore,
by analyzing whether interactions are {\em deep} or {\em shallow}, we can determine if users engage more deeply with certain content types (e.g., fake news) in private or public spaces. This distinction is crucial, as deeper engagement—such as sharing or commenting—can significantly enhance an article's impact compared to shallow actions like pressing Facebook’s ``like" button~\cite{reutersReport2023}.

{\bf Research questions and methodology:} 
Guided by the goal of better understanding how representative the public sphere of Facebook's ecosystem (typically studied in prior works) is of the private sphere and the ecosystem as a whole,
we designed our study to address the following research questions:
\begin{enumerate}
\item [\textbf{RQ1}]  How do the patterns of public and private
sharing of news articles on Facebook differ for articles with 
varying levels of 
bias and reliability?
\item [\textbf{RQ2}]  
If and how does the depth of user interactions with news articles differ within public and private sharing contexts for articles with 
varying levels of
bias and reliability?
\end{enumerate}

 We took several steps to address these previously unaddressed questions in as controlled matter as possible.  First, in contrast to most prior works that use publisher-level labeling 
 (e.g., Adelson et al. \cite{edelson2021understanding}; Horne et al. \cite{horne2019different}), we employ article-level labeling of bias and reliability.  By doing so, we capture that not all articles by a publisher have the same bias or reliability, recognizing that the publisher (source) is only one factor~\cite{understanding_how_readers} in the bias/reliability.
In particular, we consider the bias and reliability of articles ($N=19,050$), meticulously selected from a substantial pool of publishers ($N=1,121$) and 30K+ manually labeled news articles. Here, the original labeling was provided by Ad Fontes Media~\cite{AdFontes}, 
which provides bias and reliability of news articles.
Second, through active back-and-forth discussions and feedback with the Crowdtangle team, we developed and implemented a methodology using their Chrome addon that allows us to obtain simultaneous interaction statistics for a representative sample of publicly shared Facebook posts linking these articles, as well as across all Facebook posts (including both public and private posts) 
linking these articles.  
Finally, using this unique dataset, we perform a comparison of interaction dynamics when users share news articles with different levels of bias and reliability, both publicly and privately.

 {\bf Empirical example findings:}
Our analysis reveals several key insights into how users interact with news articles of varying bias and reliability on Facebook. For instance, we show that users tend to engage more deeply in private discussions than in public ones, irrespective of the news class.
When considering the news class, we highlight that users exhibit 
relatively higher deep interaction levels
with highly-unreliable content. Our results also
show that reliable news content has significantly lower private interaction shares compared to the highly-reliable or even unreliable content. 

 \textbf{Example beneficiaries}: 
Our methodology and findings contribute to a better understanding of 
news sharing and interaction dynamics on Facebook,
offering valuable insights for various stakeholders.
Media researchers and journalists can use our analysis of how bias and reliability shape engagement in public and private spheres for improved content creation and distribution strategies. 
For content moderators,
the analysis offers data-driven guidance on 
prioritizing efforts to curb the spread of problematic content. 
For policymakers, we highlight how privacy settings influence engagement patterns, with highly unreliable news, for example, garnering more engagement in the private sphere. 
These results underscore how platform design choices can influence selective exposure and engagement, potentially exacerbating issues like ideological polarization and misinformation spread.

 \textbf{Roadmap:} 
Sect.~\ref{sec:method} explains our 
research design, including news article selection, labeling, 
data collection, and the processing steps taken to ensure fair comparisons.
The resulting dataset is summarized in Sect.~\ref{sec:dataset}. Sect.~\ref{sec:private_vs_public} analyzes public vs. private dynamics both from a high-level aggregated perspective (Sect.~\ref{sec:aggregated_analysis_public_interaction_share}) and then using a detailed statistical analysis (Sect.~\ref{sec:granular_statistical_analysis}). In Sect.~\ref{sec:deep_vs_shallow_interactions}, we turn our attention to differences in the depth of interactions.
\CRedit{}{Topic and top-publishers analysis are provided in 
Sects.~\ref{sec:topic_analysis} and~\ref{sec:top_publishers}.}   
Finally, related works (Sect.~\ref{sec:related_works}) and conclusions (Sect.~\ref{sec:conclusions}) are presented.

 \section{Research Design, Data Collection and Limitations}
\label{sec:method}

At a high level, our research design has three parts.

{\bf Part 1. Article selection and labeling:}
We first obtained bias and reliability scores for the articles evaluated by Ad Fontes Media~\cite{AdFontes}. Subsequently, each article was categorized into one of three bias classes and one of four reliability classes based on these scores.

{\bf Part 2. Collection of interactive data:}
After 
 careful preprocessing of the 
URLs, we used 
the CrowdTangle browser extension to obtain interaction data for two sets of posts linked to these URLs (1)  {\em public} and (2) {\em combined} (public + private).
To address some limitations of the 
API and  to ensure fair comparison of the sets, we leave a four-month gap between the latest labeled article (Nov. 1, 2022) and the primary data collection (Mar. 2023), as well as 
 apply 
some additional post-processing (e.g., examining the actual posts), filtering (based on thresholds determined via 
discussions
with the CrowdTangle team), and collect some complementing data directly from CrowdTangle (to address limitations of the extension API).

{\bf Part 3. Data analysis:}
Finally, we use the
 final dataset (capturing the two sets) 
to compare the properties of the {\em public} vs. {\em combined} (private + public) sets and identify significant statistical differences between the interaction patterns of public and private posts.  
We next provide details of the initial two steps, before presenting our 
analysis results (Part 3) in the subsequent sections.

\subsection{News Article Selection and Labeling}
\label{sec:collection-labeling}

{\bf Selection of articles:}
There are several independent initiatives that assess the bias and/or reliability of individual news articles and/or news sources. Examples of such evaluation efforts include Media Bias Fact Check~\cite{MBFC}, Ad Fontes Media, AllSides~\cite{AllSides}, and NewsGuard~\cite{NewsGuard}. Among these, we opted to use data from Ad Fontes Media for the following main reasons: 
 (1) it has been widely used in 
previous research~\cite{haq2022s,huszar2022algorithmic,knutson2024news}, (2) they evaluate individual news articles (not only the news publishers), (3) each assessed article receives a score for both bias and reliability, (4) they offer a transparent evaluation methodology, which is published and explained in a white paper~\cite{2021AdPaper}, and finally (5) the dataset comprises a large set of news contents 
randomly selected from a diverse set of publishers.

We obtained all evaluated news from Ad Fontes Media as of Nov. 1, 2022.  
This 
included 31,446 news article URLs from 1,121 publishers.
After filtering out URLs
serving
as event reporting pages (e.g., \url{https://www.nola.com/news/hurricane}) and articles with updated ratings (per Ad Fontes Media’s documentation), the final dataset included 31,408 articles.

{\bf Labeling of articles:}
Each article in the Ad Fontes Media dataset is assessed for both bias and reliability by a minimum of three human analysts, representing a mix of right, left, and center self-reported political perspectives. Bias scores from Ad Fontes Media range between -42 and +42, 
where more negative values suggest a stronger left-ward bias, and positive values indicate a right-leaning bias.
As for reliability, scores vary from 0 to 64, with 64 representing the most-reliable news.

For our analysis, we categorized the bias and reliability scores into distinct classes. While previous work mainly focuses on binary classification of news
 (e.g., Left vs. Right or Fake vs. True), we opted for a more granular approach by defining three classes of bias (\textit{Left}, \textit{Right}, and \textit{Center}) and four classes of reliability (\textit{Most-unreliable}, \textit{Unreliable}, \textit{Reliable}, and \textit{Most-reliable}). 
Figures~\ref{fig:bias_cdf} and~\ref{fig:reliability_cdf} show the Cumulative Distribution Function (CDF) of the bias and reliability scores of the articles included in our analysis (after additional filtering steps, explained later in this section) together with the threshold values and class labels used to define each class.

\aptLtoX{
\begin{figure}[t]
\centering
\begin{subfigure}{0.48\linewidth}
    \centering
    \includegraphics[trim = 0mm 10mm 0mm 0mm, width=0.98\linewidth]{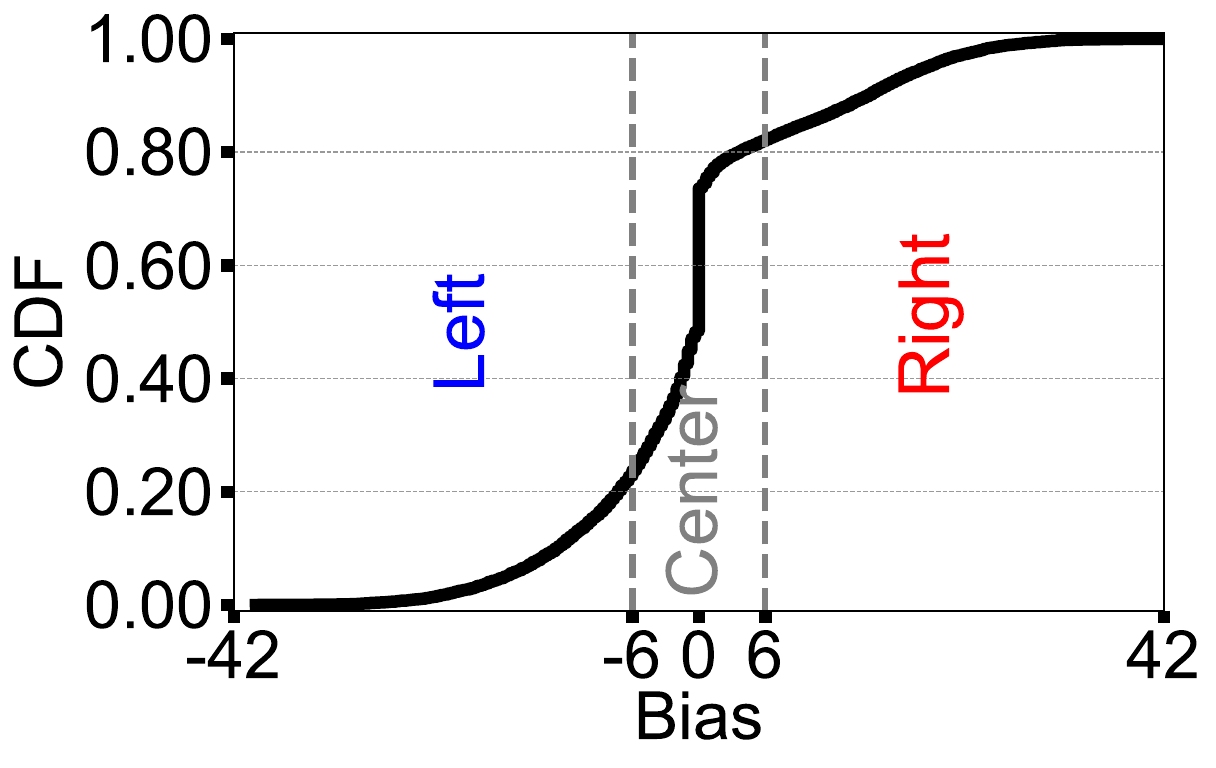}
    \caption{Bias scores}
    \label{fig:bias_cdf}
\end{subfigure}%
\hfill
\begin{subfigure}{0.48\linewidth}
    \centering
    \includegraphics[trim = 0mm 10mm 0mm 0mm, width=0.98\linewidth]{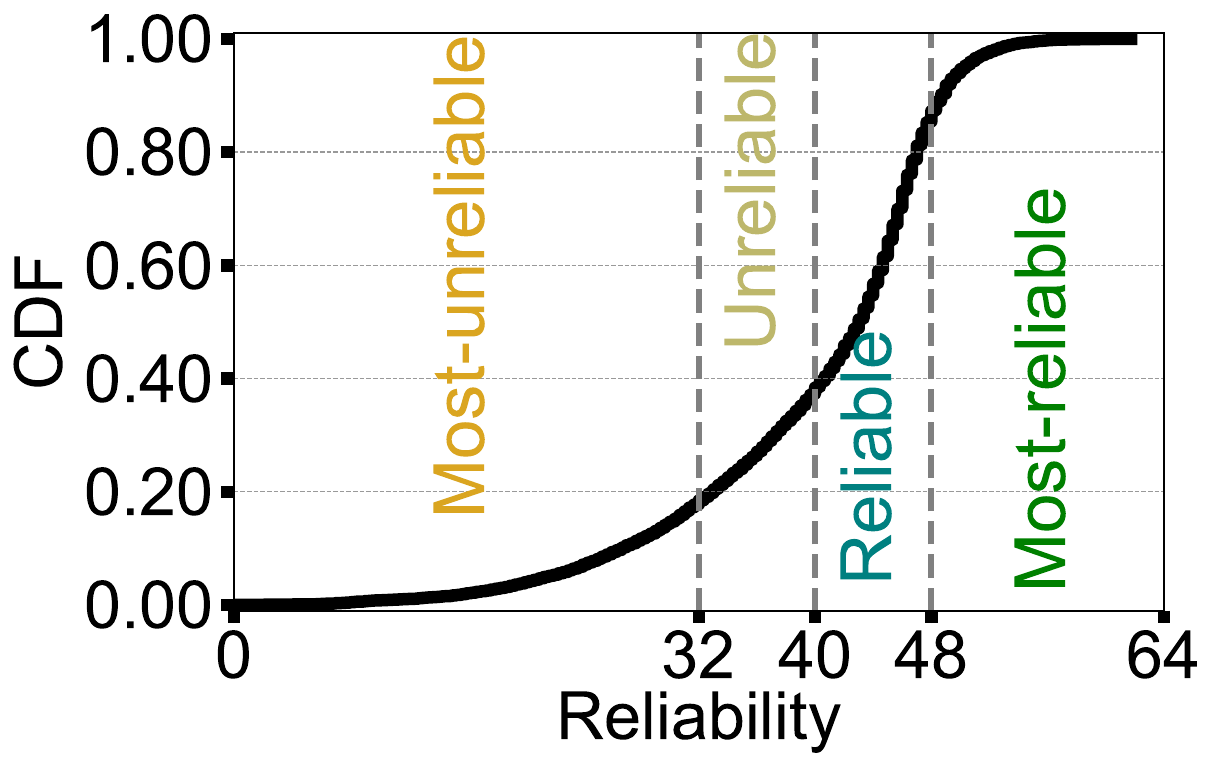}
    \caption{Reliability scores}
    \label{fig:reliability_cdf}
\end{subfigure}
\vspace{-6pt}
\caption{CDFs of scores for (a) bias and (b) reliability.}
\label{fig:combined_cdf}
\end{figure}}{}

With our labeling, anything with a bias score below -6 is called {\em Left}-biased, and anything above 6 is called {\em Right}-biased, with the range (-6,6)  capturing the \textit{Center} class. 
We note that Ad Fontes Media also refers to this range as ``Middle or Balanced Bias''~\cite{AdFontes}. As observed, 
this class represents most
samples in our dataset, representing 58\% (11,094 out of 19,050 articles in the final dataset). With this split, the two biased classes have 
similar, close to 
 20\%, shares.
 
For the reliability classes, we assigned the \textit{Most-unreliable} class to cover the lowest range (0, 32).
According to Ad Fontes Media's terminology~\cite{AdFontes}, these values typically correspond to articles ``which contain inaccurate and misleading info, selective stories or have opinion and wide variation in their reliability''. Similarly, the range (48-64), chosen for the \textit{Most-reliable} class, encompasses articles characterized by ``thorough and original fact reporting'' due to Ad Fontes terminology.  
Finally, the threshold of 40, between \textit{Reliable} and \textit{Unreliable}, matches the mid-point between these two classes as well as where Ad Fontes make their split between ``wide variation in reliability" and ``mix of fact reporting and analysis".

\subsection{Collection of Interaction Data}

{\bf Preprocessing of URLs:}
Before collecting the interaction data associated with each article (URL), we 
(1) expanded link shorteners and (2) converted URLs to canonical forms.
During the conversion process, we primarily 
removed unnecessary URL parameters after ``?" except those essential to canonical forms (e.g., ``id" parameters are sometimes part of the canonical form).
We also replaced web archive links in the Ad Fontes Media’s dataset with original links.

{\bf Primary data collection for the two sets of posts:}
Prior to its shutdown 
\CRedit{by Facebook in}{in} 
Aug. 2024, CrowdTangle (owned by Facebook) provided access to interaction statistics for public posts, which it indexed in a database widely used by researchers. Additionally, their Chrome extension offered access to the life-time interaction statistics available at the time, encompassing all posts (public + private). In this study, we utilized the extension to collect statistics for two datasets: \textit{combined} and \textit{public}, where the {\em combined} dataset includes all interactions from all posts (public + private), and the {\em public} dataset represents a subset of the combined dataset, containing only interactions with publicly accessible posts.

\aptLtoX{}{
\begin{figure}[t]
\centering
\begin{subfigure}{0.48\linewidth}
    \centering
    \includegraphics[trim = 0mm 10mm 0mm 0mm, width=0.98\linewidth]{bias_cdf.pdf}
    \caption{Bias scores}
    \label{fig:bias_cdf}
\end{subfigure}%
\hfill
\begin{subfigure}{0.48\linewidth}
    \centering
    \includegraphics[trim = 0mm 10mm 0mm 0mm, width=0.98\linewidth]{reliability_cdf.pdf}
    \caption{Reliability scores}
    \label{fig:reliability_cdf}
\end{subfigure}
\vspace{-6pt}
\caption{CDFs of scores for (a) bias and (b) reliability.}
\label{fig:combined_cdf}
\vskip-10pt
\end{figure}}{}

To collect the data for each URL, we used the ``Download" option within the extension's interface when browsing each article.  
The downloaded CSV file contained
the interaction data for both sets. As expected from a privacy standpoint, only the aggregated statistics are provided for the \textit{combined} set. In contrast, every post and its individual interaction data are provided for the public set.

 At a high level, the 
\textit{combined} category aggregates the interactions (e.g., likes, shares, comments, etc.) of all Facebook posts referencing the 
article's URL, regardless of the post's privacy settings. This includes interactions on posts 
with
limited privacy settings, such as ``Friends only" or ``Only Me". In contrast, the \textit{public} category, 
 includes all interactions with the public posts tracked by CrowdTangle, which, due to heavy-tailed characteristics and significant coverage, captures most interactions with public Facebook posts. As supporting examples, all posts of all US-based public groups with 2K+ members are indexed, and in 2021 over 99\% of all Facebook pages with at least 25K likes were indexed~\cite{crowd-tangle-tracking}.

{\bf Timeline and limitations:}
We 
collected the interaction data in Mar. 2023, ensuring at least a four-month gap between the publication date of the newest article in our dataset and our interaction data collection. This interval ensured capturing most posts and interactions for the URLs. 
This four-month window is notably conservative, as research has shown that most interactions occur within days of publication~\cite{elin,half_life_of_tweet}.
\CRedit{}{Although this conservative window may have overlooked interactions with content that was later removed or deleted, we are unable to assess the extent of this effect—if it exists—due to CrowdTangle’s shutdown in Aug. 2024.}

As 
mentioned in~\cite{cr-addon-why}, the information from the CrowdTangle addon 
originated
from two different sources: (1) public interaction data is obtained from the CrowdTangle database, while (2) combined interaction data is derived from Facebook's Graph API. Consequently, variations in data capturing times might cause some URLs to exhibit higher public than combined interactions.

Furthermore, 
as confirmed by the CrowdTangle team,
Facebook's aggregation data restarts the interaction count whenever websites update their connection schemes (e.g., switching from HTTP to HTTPS) after publishing a URL, while CrowdTangle 
continued
accumulating data.
Although the discrepancy is not substantial (9\% in our initial dataset), we aimed to mitigate the effect of these differences where possible (described next).

\aptLtoX[graphic=no,type=html]{\begin{table}[t]
\centering
  \centering
  \captionof{table}{Statistics based on Bias and Reliability.}
  \label{tab:bias_reliability_classes_stats}
  \vspace{-6pt}
  \small
  \begin{tabular}{l@{\hspace{2em}}cccc}
    \toprule
    & Class & Articles no & Combined interactions & Public interactions \\
    \midrule
    \multirow{-3}{*}{\raisebox{-3.5em}{\rotatebox[origin=c]{90}{Bias}}}
    & Left & 4,516 & 79,038,246 & 20,307,107 \\
    & Center & 11,094 & 90,084,398 & 32,109,924 \\
    & Right & 3,440 & 56,160,101 & 18,011,285 \\
    \hline
    \multirow{-4}{*}{\raisebox{-5.5em}{\rotatebox[origin=c]{90}{Reliability}}}
    & Most-unreliable & 3,520 & 45,299,475 & 13,206,643 \\
    & Unreliable & 3,795 & 46,712,927 & 14,199,322 \\
    & Reliable & 9,308 & 89,596,137 & 30,639,973 \\
    & Most-reliable & 2,427 & 43,674,206 & 12,382,378 \\
    \hline
    & \textbf{Total} & \textbf{19,050} & \textbf{225,282,745} & \textbf{70,428,316} \\
    \bottomrule
  \end{tabular}
\end{table}%
\begin{figure}
  \centering
  \includegraphics[width=\linewidth]{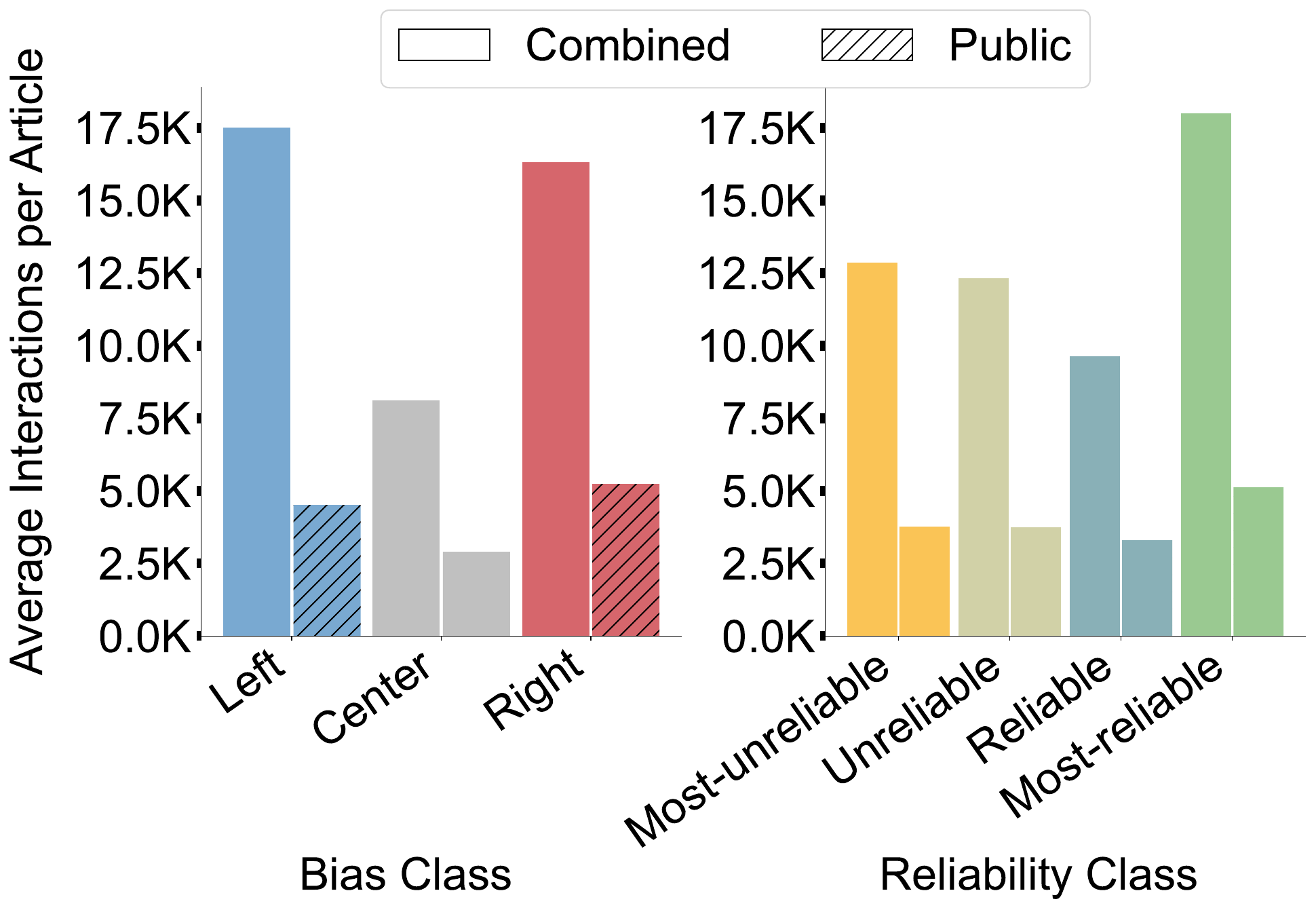}
  \vspace{-14pt}
  \caption{Interactions per article}
\label{fig:aggregated_average_interactions_per_article}
\end{figure}
}{\begin{figure*}[t]
\centering
\begin{minipage}[c]{0.6\textwidth}
  \centering
  \captionof{table}{Statistics based on Bias and Reliability.}
  \label{tab:bias_reliability_classes_stats}
  \vspace{-6pt}
  \small
  \begin{tabular}{l@{\hspace{2em}}cccc}
    \toprule
    & Class & Articles no & Combined interactions & Public interactions \\
    \midrule
    \multirow{-3}{*}{\raisebox{-3.5em}{\rotatebox[origin=c]{90}{Bias}}}
    & Left & 4,516 & 79,038,246 & 20,307,107 \\
    & Center & 11,094 & 90,084,398 & 32,109,924 \\
    & Right & 3,440 & 56,160,101 & 18,011,285 \\
    \midrule
    \multirow{-4}{*}{\raisebox{-5.5em}{\rotatebox[origin=c]{90}{Reliability}}}
    & Most-unreliable & 3,520 & 45,299,475 & 13,206,643 \\
    & Unreliable & 3,795 & 46,712,927 & 14,199,322 \\
    & Reliable & 9,308 & 89,596,137 & 30,639,973 \\
    & Most-reliable & 2,427 & 43,674,206 & 12,382,378 \\
    \midrule
    & \textbf{Total} & \textbf{19,050} & \textbf{225,282,745} & \textbf{70,428,316} \\
    \bottomrule
  \end{tabular}
\end{minipage}%
\hfill
\begin{minipage}[c]{0.34\textwidth}
  \centering
  \includegraphics[width=\linewidth]{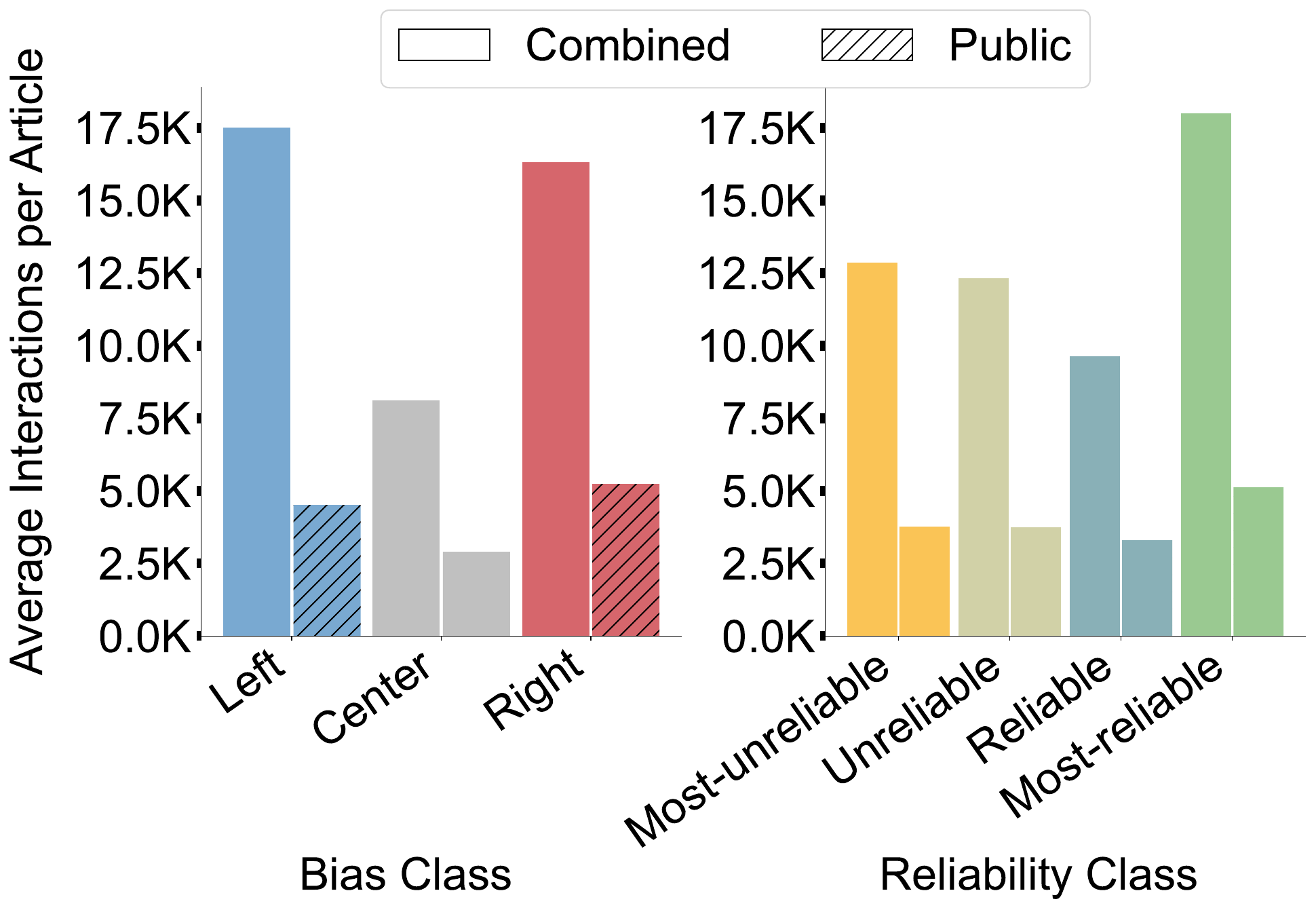}
  \vspace{-14pt}
  \caption{Interactions per article}
\label{fig:aggregated_average_interactions_per_article}
\end{minipage}
\vspace{-6pt}
\end{figure*}}

{\bf Further filtering and data processing for fairer comparisons:}
After consulting with the CrowdTangle team regarding all numerical discrepancies that we observed, we identified two additional reasons for the discrepancies. First, 
as stated in \cite{facebook2022url}, for privacy reasons, for the combined data (that include private posts) ``the values are intentionally not precise, but you can be confident they accurately reflect user engagement with a URL." 
Discussing the identified discrepancy for such cases, the CrowdTangle team confirmed that combined interaction data with counts below 100 is not reliable; however, that data with counts above 100 provides a reliable estimate of public + private interactions. For this reason, we removed all URLs in our dataset with combined interactions below 100. Although this reduced our dataset to 19,505 articles, it is important to note that this group of URLs accounted for less than 4\% of public interactions in our dataset. Furthermore, we note that the public interaction data for this group still remains reliable.

Second, 
the CrowdTangle team informed us that the Facebook Graph API only considers posts with the URL attached to the post (i.e., those displaying a preview of the URL) when counting interaction with the URL, while CrowdTangle's algorithm includes any post containing a linked URL, regardless of attachment status. Taking this into account, in our final step toward enhancing interaction data quality, we chose not to rely on the aggregated data from the CrowdTangle extension (which sums up the interactions of all posts mentioned in the extension). Instead, we examined each post to determine whether the related URL was attached or not, and if so, we included them in the sum of public interactions for that URL.

This resulted in a more accurate comparison between combined and public interactions. After these processing steps, we had only 455 URLs for which public interactions exceeded combined interactions, which were excluded from our final dataset. The primary causes of these discrepancies were beyond the scope of our research to address (e.g., changes in URL, publisher protocol schemes, or different data capturing times as mentioned above).

{\bf Enhancing the data with additional posts:}
Another limitation of the CrowdTangle extension 
was its restriction to retrieving
data for up to 500 posts per URL. 
To address this, we retrieved additional data from CrowdTangle for URLs with over 500 posts in the \textit{public} interaction category. 
While some of the pre- and post-processing steps described above require significant effort, they help ensure an accurate and fair comparison between the public and combined sets in such a way that we can provide conclusive insights into the relative sharing patterns of public vs. private posts.

\section{Dataset}\label{sec:dataset}

{\bf High-level summary:}
After applying the aforementioned filtering steps, we have in total 19,050 articles from 1,121 news outlets remaining in our dataset. These articles have been shared in 253,350 posts,
which combined are responsible for 225,282,745 interactions (out of which 70,428,316 are public).  

{\bf Bias data:}
Table~\ref{tab:bias_reliability_classes_stats}
shows the number of articles and the total combined and public interactions for each bias and reliability class.
With the selected bias thresholds, our dataset includes 4,516 \textit{Left}-biased articles and 3,440 \textit{Right}-biased articles, with the remaining 11,094 articles falling in the \textit{Center} category.
These articles are in turn responsible for 79,038,246 (\textit{Left}), 56,160,101 (\textit{Right}), and 90,084,398 (\textit{Center}) interactions.
While the \textit{Center} category includes 58\% of the articles, it is interesting to note that it is responsible for a significantly smaller fraction of the total interactions (40\%).  Instead, the \textit{Left}-biased and \textit{Right}-biased articles see relatively higher interaction rates (studied in the next section).  For example, the \textit{Left}-biased articles account for only 24\% of the articles but 35\% of the interactions, and the \textit{Right}-biased articles are responsible for only 18\% of the articles but 25\% of the interactions.

{\bf Reliability data:}
For the reliability classes as shown in
  Table~\ref{tab:bias_reliability_classes_stats}, 
the relative differences are smaller.
Here, the \textit{Reliable} articles make up the largest share (9,308 articles and 89,596,137 interactions), followed by \textit{Unreliable} (3,795 articles and 46,712,927 interactions), \textit{Most-unreliable} (3,520 articles and 45,299,475 interactions), and \textit{Most-reliable} (2,427 articles and 43,674,206 interactions).

\section{Public vs. Combined Interactions}\label{sec:private_vs_public}
To understand how engagement patterns differ between public and private spheres on Facebook, we present two complementary analyses: (1) an aggregated comparison of public vs. combined (public + private) interactions (Sec.~\ref{sec:aggregated_analysis_public_interaction_share}), and (2) a statistical analysis of their differences (Sec.~\ref{sec:granular_statistical_analysis}).
By comparing public interactions against combined interactions, where private interactions constitute the majority (as we show later), we can effectively study how engagement in private spaces differs from public ones.

 \subsection{Aggregated (Macro) Analysis of Interactions} \label{sec:aggregated_analysis_public_interaction_share}

Figure~\ref{fig:aggregated_average_interactions_per_article} presents the 
number of combined and public interactions per article for each class. 
Here, two metrics are used:

{\bf Average combined interactions:} (left bars for each class) represents the average number of combined (private+public) interactions per article in each class.
More specifically, considering $I_{c,i}^{\text{comb}}$ as the combined interaction for the $i^{th}$ article in class $c$, the average combined interactions for this class ($\bar{I}_c^{\text{comb}}$) is calculated as:
    \begin{equation}
    \bar{I}_c^{\text{comb}} = \frac{\sum_{i=1}^{N_c} I_{c,i}^{\text{comb}}}{N_c},
    \end{equation}
where the numerator ($\sum_{i=1}^{N_c} I_{c,i}^{\text{comb}}$) corresponds to the ``Combined interactions'' column in Table \ref{tab:bias_reliability_classes_stats} and the denominator (i.e., the total number of articles $N_c$ for class $c$) is found in the ``Articles no'' column of Table \ref{tab:bias_reliability_classes_stats}.
For example, for the \textit{Center} class in the \textit{Bias} category, the average combined interactions per article is computed by dividing the total interactions (90,084,398) by the articles count (11,094), yielding $\bar{I}_{\text{Center}}^{\text{comb}}$ to be 8,120 combined interactions per article (gray bar in Figure \ref{fig:aggregated_average_interactions_per_article}).

{\bf Average public interactions per article:} (right bars) is 
    defined in a similar manner but considering only the public interactions.
    More specifically, considering $I_{c,i}^{\text{public}}$ as the public interactions for the $i^{th}$ article in a class $c$, the average public interactions for this class ($\bar{I}_c^{\text{public}}$) is calculated as follows:
    \begin{equation}
    \bar{I}_c^{\text{public}} = \frac{\sum_{i=1}^{N_c} I_{c,i}^{\text{public}}}{N_c},
    \end{equation}
where the numerator (i.e., $\sum_{i=1}^{N_c} I_{c,i}^{\text{public}}$) corresponds to the ``Public Interactions" column in Table \ref{tab:bias_reliability_classes_stats}.
Now, taking the \textit{Center} class as an example again, the average public interactions per article is calculated by dividing the total public interactions (32,109,924) by the article count (11,094), resulting in an $\bar{I}_{\text{Center}}^{\text{public}}$ of 2,894 interactions per article (striped gray bar in Figure \ref{fig:aggregated_average_interactions_per_article}).

As seen in the figure, there are some very clear and interesting differences in the trends observed for the public vs. the combined statistics.
First, for all three bias classes and for all four reliability classes, the interaction rates are significantly lower for the public posts than for the combined set.
These aggregate rate differences suggest that users are more likely to interact with private posts, regardless of bias and reliability class, and underscores the importance of considering the total (combined) interactions, not only the (typically studied) public posts. We next look closer at the relative interaction levels of each class and the impact of using the public vs. combined sets for such interaction comparisons.

{\bf Bias comparisons:}
Second, examining the relative differences
among the bias classes, the \textit{Left} bias class has the highest average combined interactions per article (17,501), followed by the \textit{Right} (16,325) and \textit{Center} (8,120) classes. 
For
public interactions; however, the \textit{Right} bias class has the highest average public interactions per article (5,236), with the \textit{Left} (4,497) and \textit{Center} (2,894) classes trailing behind.  This shows that biased articles see even greater relative interaction rates 
in private than in public posts.

{\bf Reliability comparisons:}
Third, for the reliability classes, the \textit{Most-reliable} class has the highest average combined and public interactions per article (17,995 and 5,101). The \textit{Unreliable} and \textit{Reliable} classes have relatively similar average public interactions per article (3,741 and 3,291, respectively). Moreover, the \textit{Most-unreliable} class has a higher average combined interactions per article (12,869) and public interactions per article (3,751).
It is interesting to see that the two extreme classes (i.e., the \textit{Most-reliable} and \textit{Most-unreliable}) have the highest average interactions per article in both combined and public contexts, implying that users are more likely to engage with articles on both ends of the reliability spectrum.


{\bf Aggregated public interaction share:}
Motivated by most prior works only studying the public posts tracked by CrowdTangle, we next 
\CRedit{look closer at how big}{evaluate what} 
fraction of the total interactions this set captures and compare the relative differences in 
the public interaction share of 
\CRedit{the different}{different} 
categories.
For this analysis, 
\CRedit{we calculate the (aggregate) public interaction share, defined as the ratio of public interactions to the combined interactions for each class $c$ as}{we calculate the (aggregate) public interaction share as the ratio of public interactions to total interactions for each class $c$:}
$ \bar{I}_c^{\text{public}} / \bar{I}_c^{\text{comb}}$.

\begin{figure}[t]
\centering
\includegraphics[width=0.8\columnwidth]{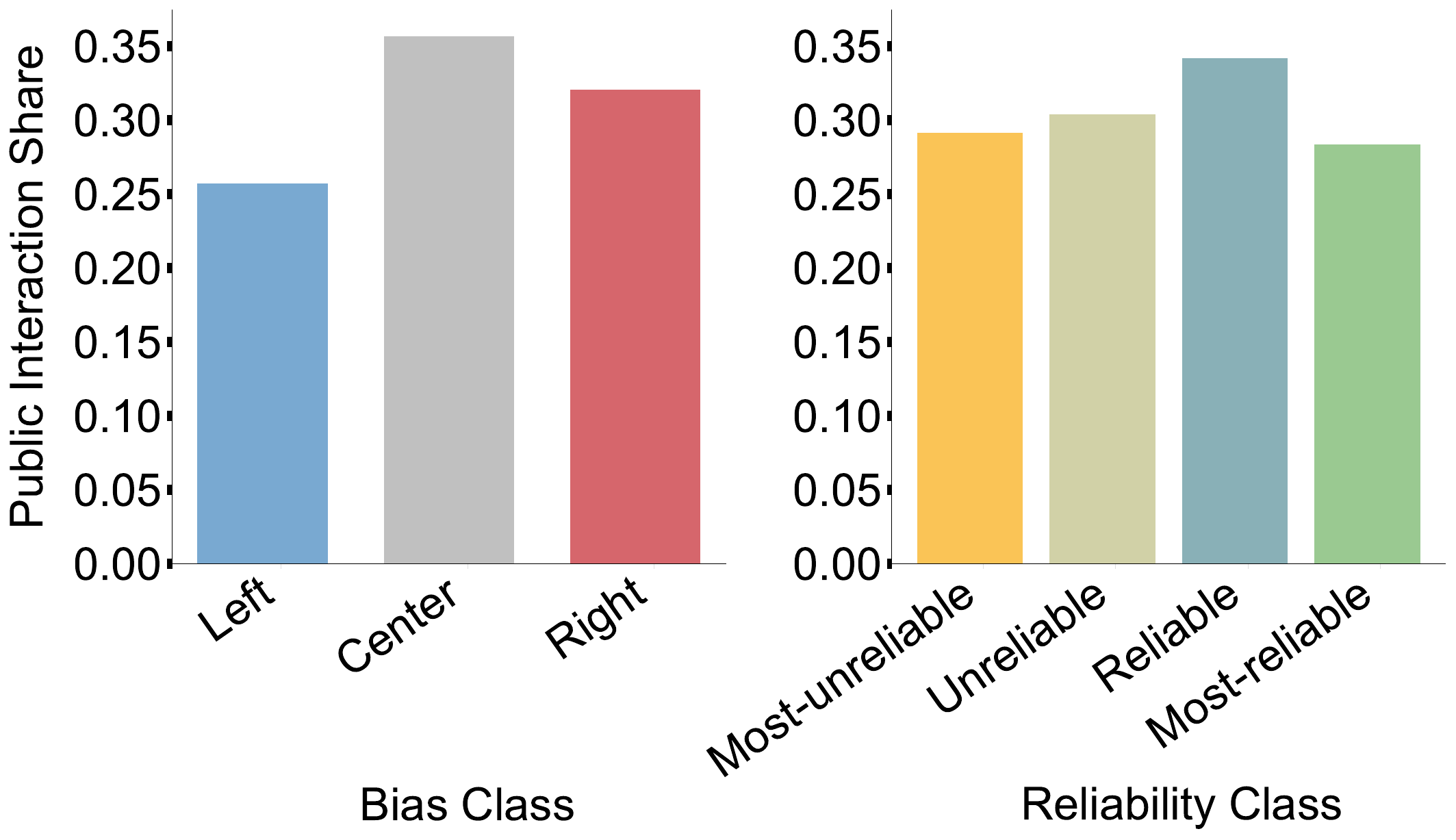}
\vspace{-10pt}
\caption{Aggregated public interactions shares}
\label{fig:aggregated_public_interaction_share}
\end{figure}

Figure~\ref{fig:aggregated_public_interaction_share} summarizes these ratios.
 We 
note that the \textit{Center} class has the highest aggregated public interaction share, and that among the biased groups, there is a clear gap between the left and right parties,
as the \textit{Right} class has a significantly (25\% extra) higher public interaction share (lower private interaction share).
When it comes to the reliability classes, the relative gap between the classes is smaller. 
However, it should be noted that the two extreme classes (\textit{Most-reliable} and \textit{Most-unreliable}) have the lowest (and almost similar) aggregated public interaction shares.
\keytakeaway{Compared to right-biased articles, left-biased articles receive higher private interactions share. In terms of reliability, extremely reliable/unreliable news articles receive higher levels of private interactions share.}

\begin{figure*}[t]
\centering
\begin{subfigure}[t]{0.48\textwidth}
    \centering
    \includegraphics[width=\linewidth]{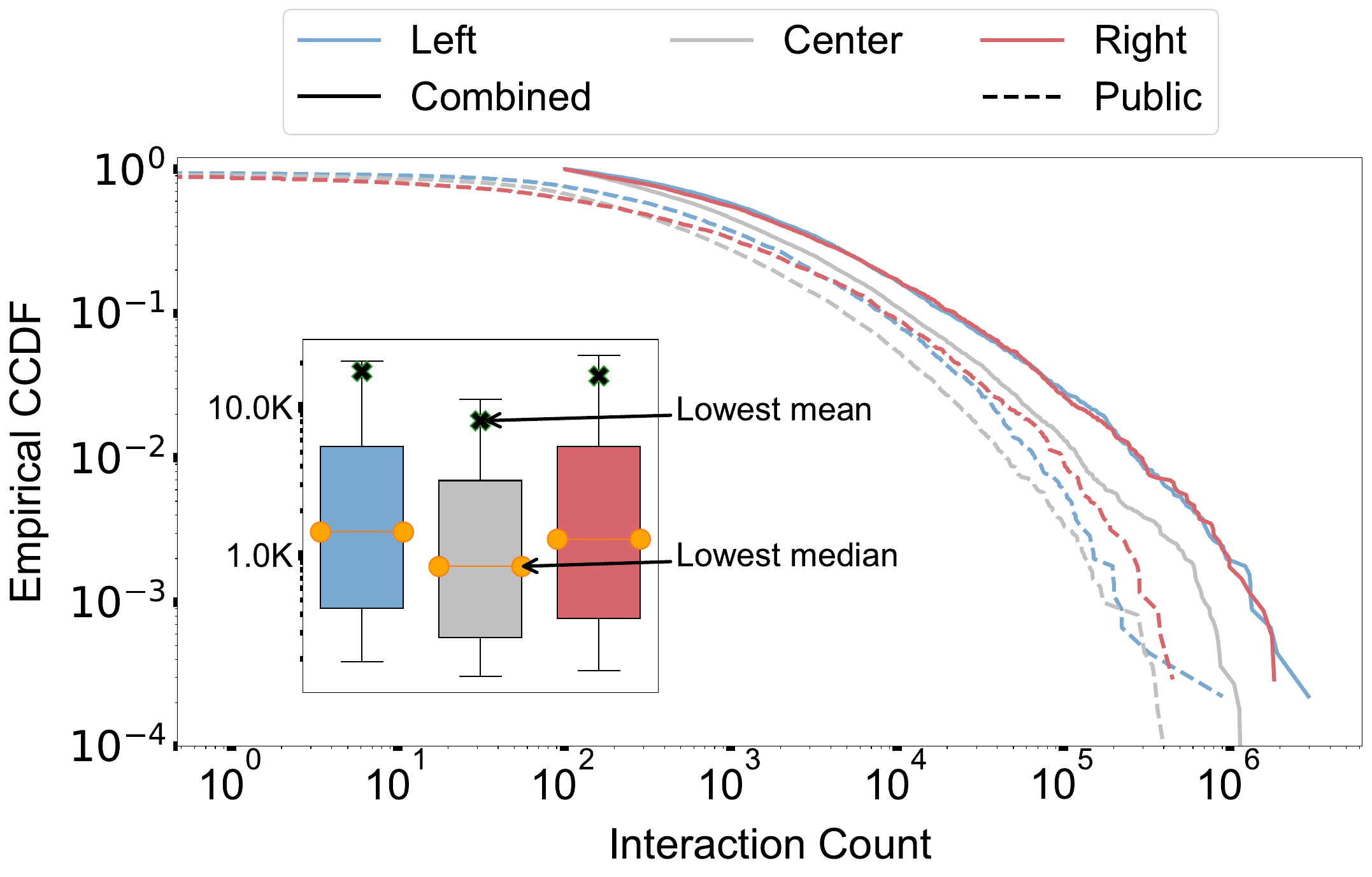}
    \caption{Disributions of different bias classes.}
    \label{fig:ccdf_of_combined_and_public_interactions_for_different_bias_classes}
\end{subfigure}%
\hspace{0.03\textwidth} 
\begin{subfigure}[t]{0.48\textwidth}
    \centering
    \includegraphics[width=\linewidth]{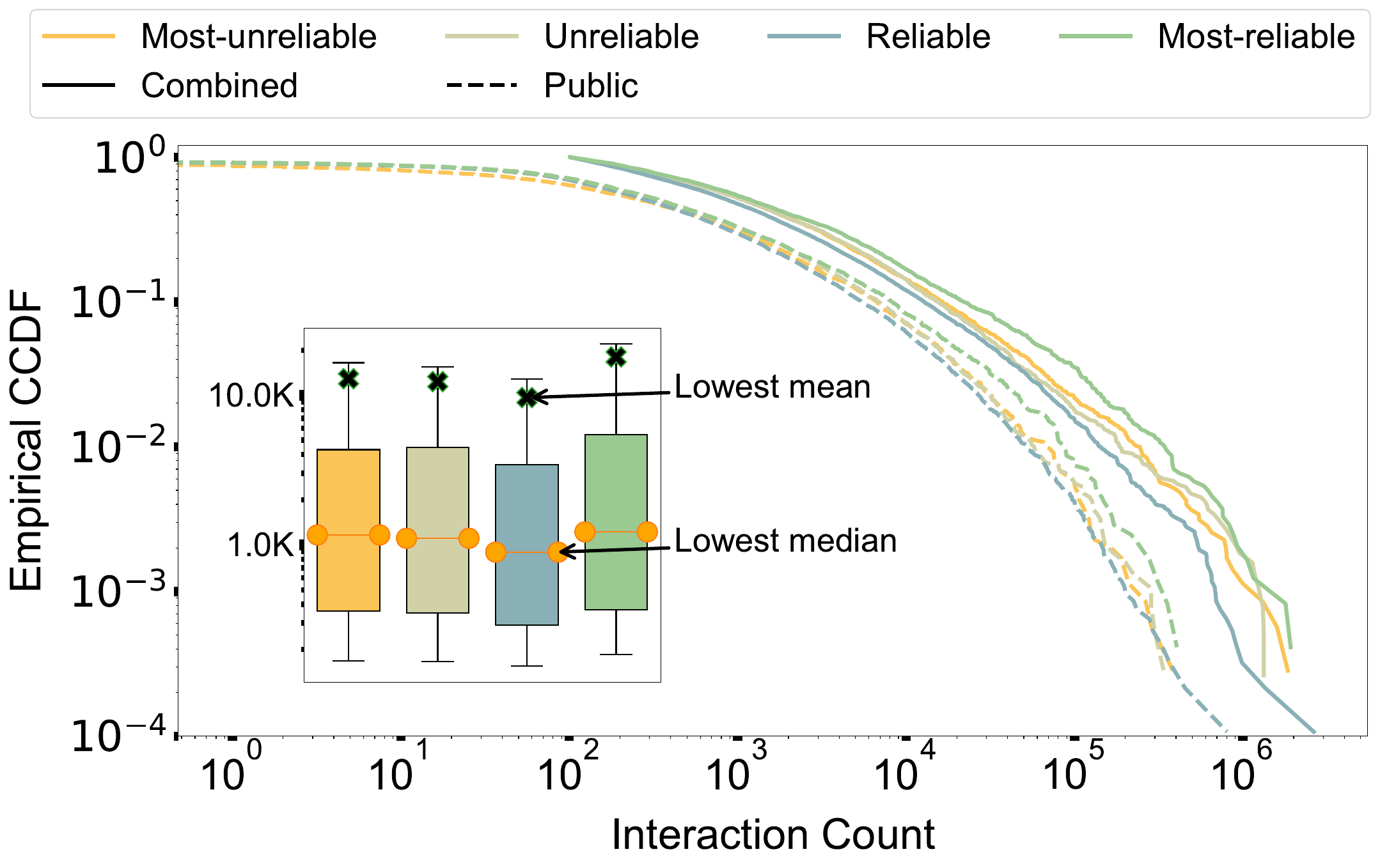}
    \caption{Disributions of different reliability classes.}
    \label{fig:ccdf_of_combined_and_public_interactions_for_different_reliability_classes}
\end{subfigure}%
\vspace{-8pt}
\caption{Empirical complementary cumulative distribution functions (CCDFs) for combined and public interactions, along with inset boxplots showing the distribution of combined interactions. Both CCDFs emphasize tail characteristics of distributions, while the boxplots provide granular statistical summaries, including medians, quartiles, and means.}
\label{fig:ccdf_combined_figure}
\end{figure*}

 \subsection{Granular Statistical (Micro) Analysis of Interactions}\label{sec:granular_statistical_analysis}
 Having presented aggregated analysis and insights, we next present a more detailed statistical comparison between different classes in which we give equal consideration to every article in a class, regardless of its total interaction count. For this analysis, we utilize the distributions of the per-article level interactions for each class $c$: $\mathcal{D}^{\text{comb}}_c = \{ I_{c,i}^{\text{comb}}\}$ and $\mathcal{D}^{\text{public}}_c = \{ I_{c,i}^{\text{public}}\}$, where $I_{c,i}^{\text{comb}}$ is the total number of interactions across all posts associated with the $i^{th}$ article of class $c$ ($1 \leq i \leq N_c$) and $I_{c,i}^{\text{public}}$ is the corresponding number of interactions with public posts.
 
 The next two subsections compare distributions of the total (combined) interactions (Sect. \ref{sec:combined_interactions_distribution}) and study the distributions of the relative public interaction shares (Sect.~\ref{sec:statistical_analysis_of_public_interactions_share}), respectively.
In both subsections, the distributions are compared using multiple statistical tests; statistical significance is reported if a p-value is below 0.001, and supporting p-values are reported for the main findings.
 \subsubsection{Distribution of Combined Interactions}\label{sec:combined_interactions_distribution}
 {\bf Comparison of Bias Classes:}
Figure~\ref{fig:ccdf_of_combined_and_public_interactions_for_different_bias_classes} shows the CCDFs of both the set of combined interactions ($\mathcal{D}^{\text{comb}}_c$) and the set of public interactions ($\mathcal{D}^{\text{public}}_c$) for each bias class $c$ on a log-log scale. Accompanying the CCDF is a boxplot, 
highlighting key statistical percentiles for the distribution of the combined interactions ($\mathcal{D}^{\text{comb}}_c$). Specifically, the 10$^{th}$ percentile (bottom marker), 25$^{th}$ percentile (bottom of box), median (middle marker), 75$^{th}$ percentile (top of box), 90$^{th}$ percentile (top marker), and the mean (i.e., $\bar{I}_c^{\text{comb}}$) (circle). To allow higher resolution, outliers are not shown in the boxplot.

 Upon examining the figure, we can make several observations.
First, as articles with fewer than 100 combined interactions were excluded, the minimum value for $\mathcal{D}^{\text{comb}}_c$ starts at 100. Second, the shape of the of the CCDFs when plotted on log-log scale underscores the ``heavy-tailed'' nature of the interactions ($\mathcal{D}^{\text{comb}}_c$ and $\mathcal{D}^{\text{public}}_c$) of all bias classes $c$. This suggests that a small subset of the articles are responsible for a most of the interactions. Third, there are discernible variations in the distributions. For example, articles classified as biased (both \textit{Left} and \textit{Right}) consistently see higher interaction levels (i.e., right-shifted curves compared to the {\em Center} class), echoing the trends we observed in the aggregated analysis.
To better understand these disparities and their implications, we next provide a more rigorous statistical analysis.

{\bf Statistical Tests for Bias-related Differences:}
To evaluate the differences in distributions of the set of combined interactions $\mathcal{D}^{\text{comb.}}_c$ across all bias classes $c$, we employed a multi-step statistical testing approach:

\textbf{1. Overall Differences:} 
We first used the Kruskal-Wallis test to identify overall differences between the distributions between the bias classes. This test revealed significant disparities among the classes, with a p-value of $1.66\cdot10^{-65}$.

\textbf{2. Pairwise Comparisons:} 
Second, we used the Dunn test to identify specific pairs of classes that exhibited differences. This test confirmed that the two biased classes \textit{Left} and \textit{Right} are different than the \textit{Center} class, but the difference between the two biased classes themselves was not significant. Here, the most significant difference were observed between the {\em Left} and {\em Center} class (i.e., $\mathcal{D}^{\text{comb}}_{\text{Left}}$ and $\mathcal{D}^{\text{comb}}_{\text{Center}}$), registering a p-value of $2.06\cdot10^{-55}$.

\textbf{3. Median Comparisons:} 
Third, we used the Mann-Whitney U test to compare the medians of the three classes:  \textit{Left} (1,440), \textit{Center} (842), and \textit{Right} (1,286). Also, here, we observed statistically significant pairwise differences when comparing each of the two biased classes (\textit{Left} and \textit{Right}) with the \textit{Center} class, but not between each other.  Here, the two significant cases obtained p-values of $8.4\cdot10^{-57}$ and $9.1\cdot10^{-28}$, respectively.

\textbf{4. Comparing Means:} 
Finally, due to the violations of assumptions intrinsic to a t-test (namely, normality and homogeneity of variances), we instead use bootstrapping with 100K iterations and a 99\% confidence interval to compare the means.
Again, the pairwise differences in the means between \textit{Left} vs. \textit{Center} and between \textit{Right} vs. \textit{Center} are statistically significant, but not between the two biased classes themselves (i.e., \textit{Left} vs. \textit{Right}).

\keytakeaway{
Statistical analysis further strengthens the observation that biased news consistently garners higher combined engagement levels per article than center-aligned news. Interestingly, there is no statistically significant difference between the \textit{Left} and \textit{Right} class.}

 {\bf Comparison of Reliability Classes:}
Figure~\ref{fig:ccdf_of_combined_and_public_interactions_for_different_reliability_classes} shows the CCDFs for the different reliability classes, broken down for both combined and public interactions, accompanied by a boxplot offering a clearer perspective on the percentile values for the combined interactions. 
 The CCDFs again demonstrate heavy-tailed distribution characteristic, and the relative shifts of the distributions are consistent with the aggregate (average values) previously discussed (and seen in Figure~\ref{fig:aggregated_average_interactions_per_article}), with the \textit{Most-reliable} class typically getting the most interactions, followed by the unreliable classes, and subsequently by the \textit{Reliable} class. While comparing the tail parts is straightforward, comparing the whole distributions requires some care.
For example, here, the core of the boxplot does not distinctly set apart the \textit{Most-reliable} from the unreliable classes. To derive robust conclusions, we, therefore, again performed a sequence of statistical tests on the combined interactions distributions.

First, the Kruskal-Wallis test was applied to discern differences among the reliability classes. While we observed statistical differences (p-value of $7.9\cdot10^{-20}$), the magnitude of distinction among reliability classes is more nuanced than in the bias classes (which had a p-value of $1.66\cdot10^{-65}$).
Further analysis employing the Dunn's test confirmed that the \textit{Reliable} class statistically diverges from the other three. The least significant p-value here, $7.63\cdot10^{-8}$, is attributed to the comparison between the \textit{Reliable} and \textit{Unreliable} classes. The most significant, on the other hand, emerges from the \textit{Reliable} versus \textit{Most-reliable} comparison with the p-value of $7.05\cdot10^{-15}$. 
Except for these distributions, none of the other pairwise distribution comparisons resulted in statistically significant differences. 
Third, the Mann-Whitney U test reinforced our findings, particularly highlighting the statistical significance when comparing the medians of the \textit{Reliable} class (median of 891) with the others: \textit{Most-unreliable} (1,162), \textit{Unreliable} (1,104), and \textit{Most-reliable} (1,218). 
Finally, we apply bootstrapping to compare the means.  While only the \textit{Reliable} and \textit{Most-reliable} classes comparison is significant at the 99\% confidence level, we note that all three comparisons against the \textit{Reliable} class are significant at the 95\% confidence level.

\keytakeaway{Comparing to the other reliability classes, the \textit{Reliable} articles demonstrate lower interaction values, especially when compared to the \textit{Most-reliable} articles.}
 \subsubsection{Distribution of Public Interaction Shares}\label{sec:statistical_analysis_of_public_interactions_share}
The preceding analysis centered on the distribution of combined interactions. Like the aggregated analysis, we next shift our focus to the proportion of interactions that are public. Specifically, for each class, we analyze the distribution of public interaction shares at the granularity of individual articles. For this analysis, we define the set of public interaction ratios (\textit{PIR}) of class $c$ as $\mathcal{D}^{\text{PIR}}_c = 
\left\{\frac{I_{c,i}^{\text{public}}}{I_{c,i}^{\text{comb}}} \right\}$,
where $1 \leq i \leq N_c$ is the index of articles included in class $c$.

\begin{figure}[t]
\centering
\includegraphics[width=0.8\columnwidth]{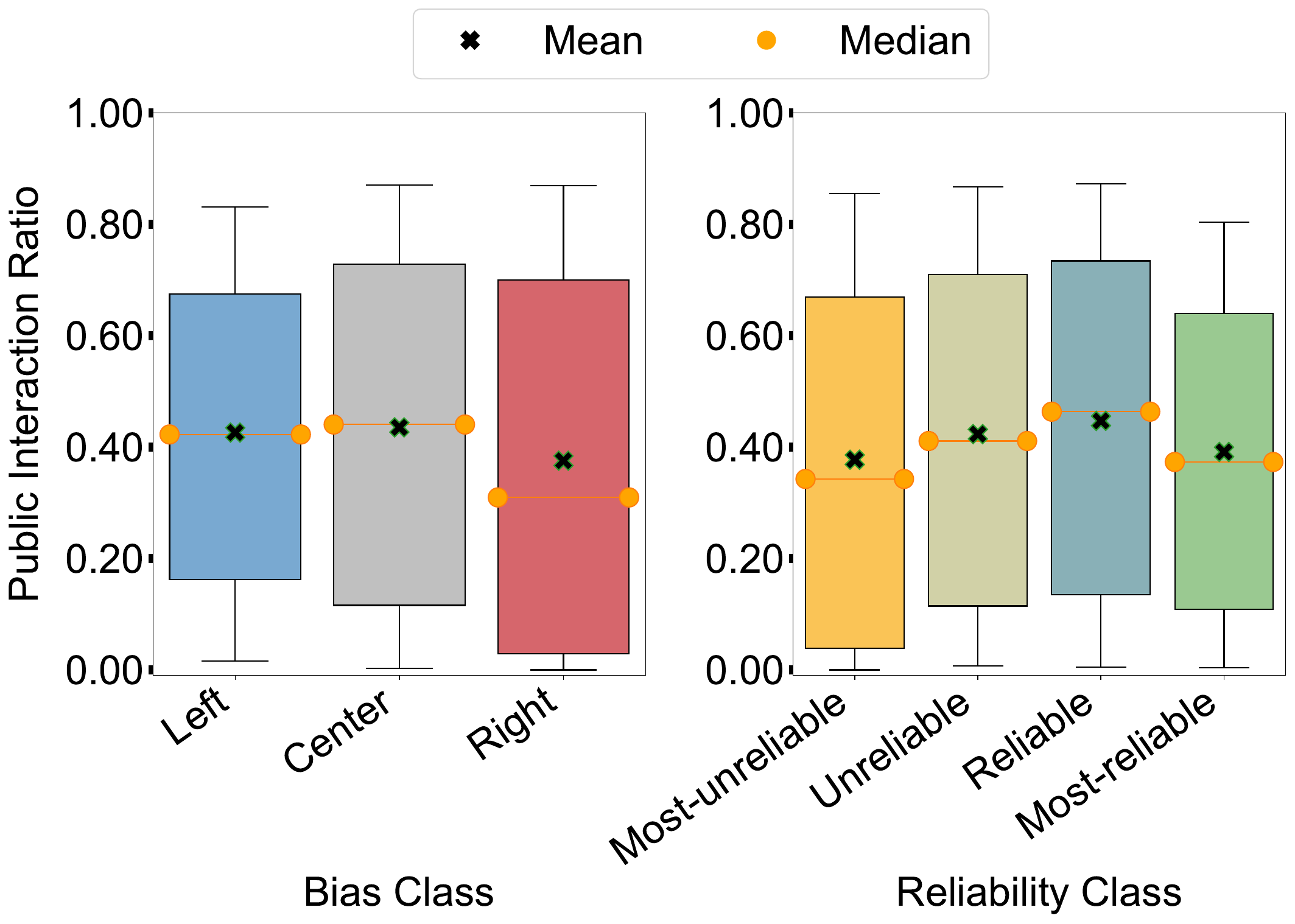} 
\vspace{-6pt}
\caption{Distribution of public interaction ratios}
\label{fig:distribution_of_public_interaction_share}
\vspace{-10pt}
\end{figure}

Figure~\ref{fig:distribution_of_public_interaction_share}
presents 
 the public interaction ratios using 
box-plots broken down per bias and reliability category. Here, we 
 again 
show the 10$^{th}$ percentile,
25$^{th}$ percentile,
median,
75$^{th}$ percentile,
and 90$^{th}$ percentile,
as well as the average.
 Like for the aggregate analysis, we observe some notable differences, which we can now support using statistical analysis. First, the Kruskal-Wallis test shows that there are statistically significant differences between 
 both the biases classes (p-value of $2.68\cdot10^{-22}$) and reliability classes (p-value of $8.74\cdot10^{-33}$).
 
Motivated by the higher significance (smaller p-value) for the reliability classes, we consider these differences first. Using the Dunn test, we find 
statistically significant differences for all pairwise
 distribution comparisons except between the \textit{Most-unreliable} and \textit{Most-reliable} classes. 
Similarly, the Mann-Whitney U test shows that all pairwise relative differences in the medians (observed in Figure~\ref{fig:distribution_of_public_interaction_share}), except for between the \textit{Most-unreliable} and \textit{Most-reliable} classes, are significant (with the largest p-value among the these pairwise cases being $2.82\cdot10^{-14}$).
Finally, bootstrapping confirms 
 that all pairwise differences in the means observed in the figure are statistically significant, except for the 
 case of the 
\textit{Most-unreliable} 
 vs. the \textit{Most-reliable} 
 class.
It is also worth noting that our observation that the two extreme classes have lower public interaction ratios (higher private interaction ratios) aligns with our findings from the aggregated analysis. 

\keytakeaway{The two extreme classes, \textit{Most-reliable} and \textit{Most-unreliable}, exhibit similar private interaction ratios that are statistically higher compared to the other classes.}

We now shift our attention to the public interaction ratios across the bias classes, where we make some interesting observations regarding the classes’ relative order.  Specifically, we now observe the first difference in the relative order of the classes when comparing the average (and median) public shares calculated on a per-article basis (Figure~\ref{fig:distribution_of_public_interaction_share}) with the aggregated results (Figure~\ref{fig:aggregated_public_interaction_share}). More specifically, comparing with the aggregate results, 
the {\em Right} class has switched ranking with the {\em Left} class, becoming the one with smallest
public interaction share.  Furthermore, comparing the distributions seen for the {\em Right} and {\em Left} class is statistically significant: Dunn test (p-value of $3.25 \times 10^{-14}$), Mann-Whitney U test (p-value of $6.16 \times 10^{-17}$) and bootstrap results being significant at 99\% level.

\aptLtoX[graphic=no,type=html]{\begin{figure}[t]
\begin{minipage}{.32\textwidth}
	\includegraphics[width=0.98\linewidth]{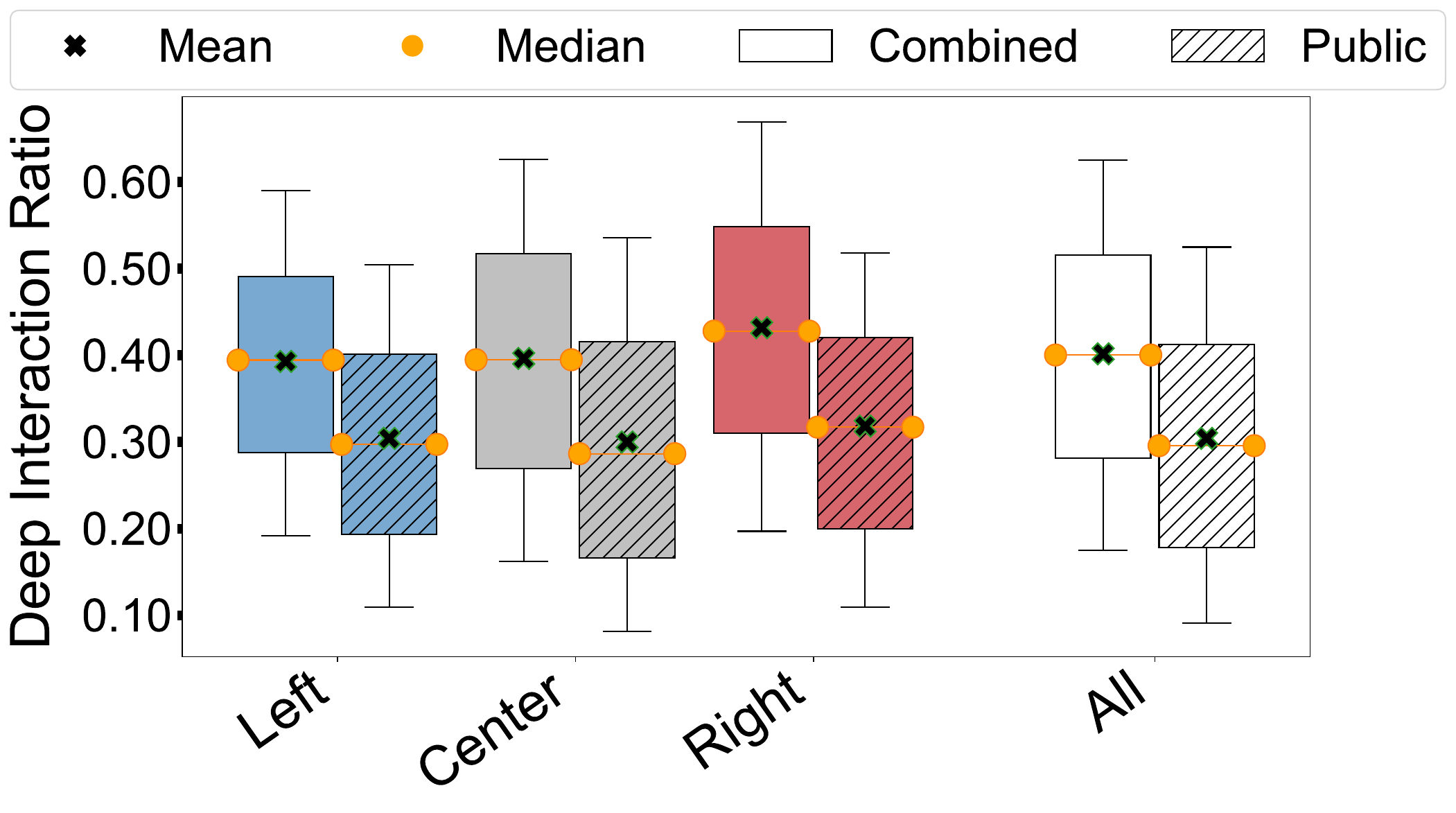}
	\vspace{-8pt}
	\caption{Deep interaction ratios (\textit{DIRC} and \textit{DIRP}) for different bias classes and a baseline.}
	\label{fig:deep_interaction_ratio_bias_classes}
\end{minipage}
\end{figure}
\begin{figure}
\begin{minipage}{.32\textwidth}
    \centering
	\includegraphics[trim =12mm 0mm 4mm 0mm, width=0.98\linewidth]{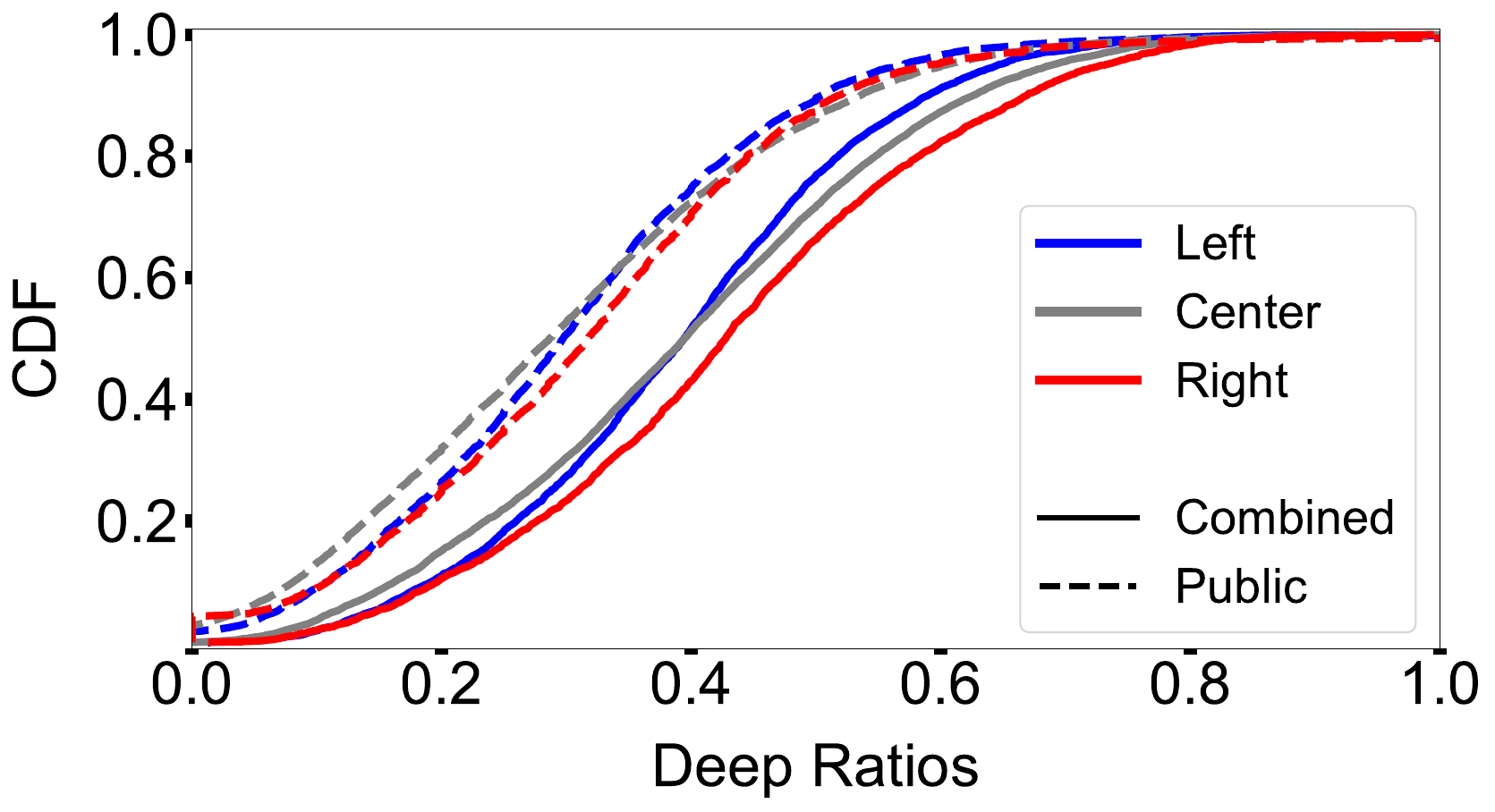} 
    \vspace{-8pt}
	\caption{CDFs of \textit{DIRC} (combined) and \textit{DIRP} (public) for different bias classes.}
	\label{fig:cdf_of_both_deep_ratio_bias_classes}
    \Description{...}
\end{minipage}
\end{figure}
\begin{figure}
\begin{minipage}{.32\textwidth}
	\centering
	\includegraphics[trim =16mm 0mm 0mm 0mm,width=0.98\linewidth]{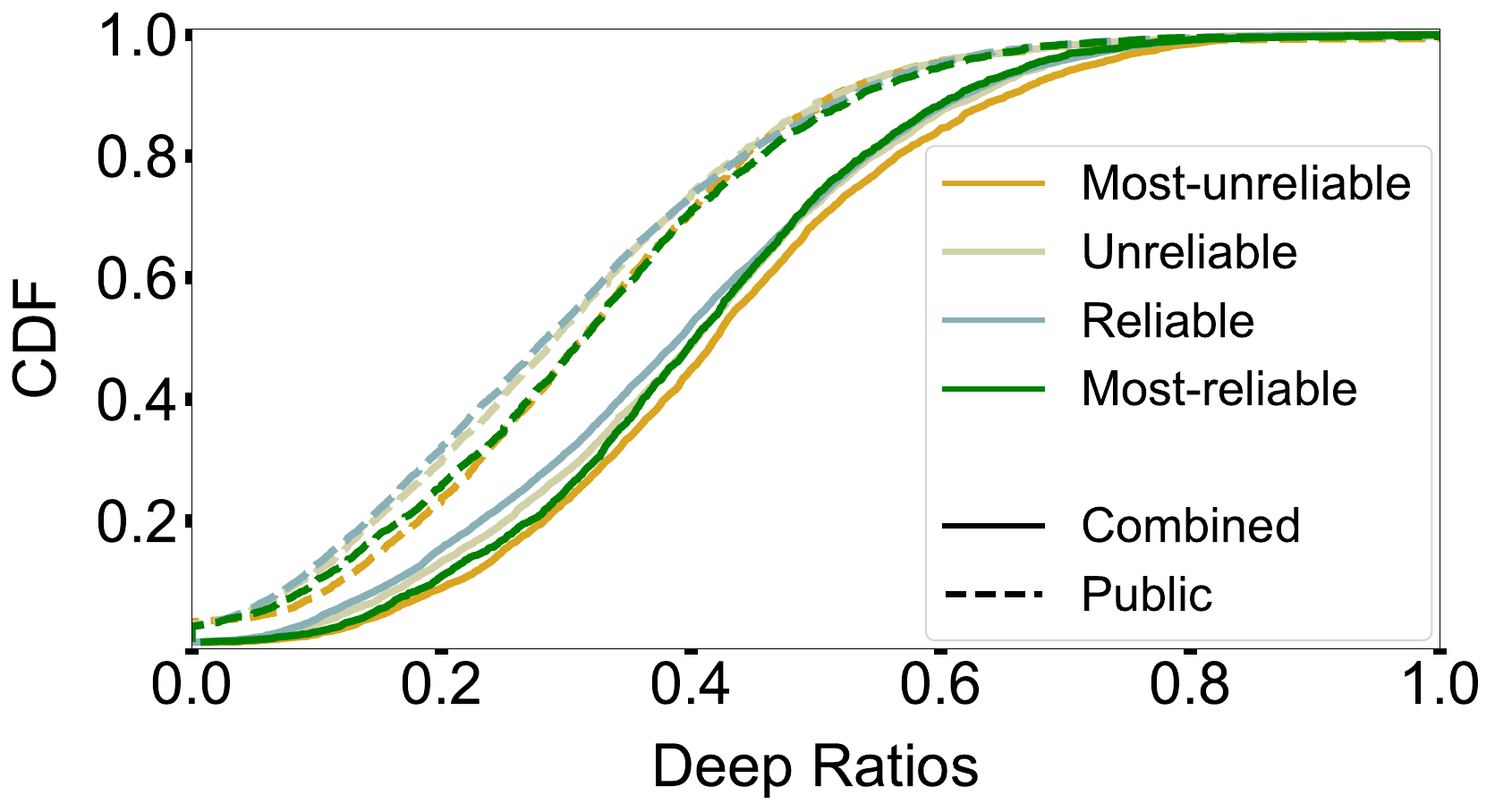} 
	\vspace{-8pt}
    \caption{CDFs of \textit{DIRC} (combined) and \textit{DIRP} (public) for different reliability classes.}
    \label{fig:cdf_of_both_deep_ratio_reliability_classes}
	\Description{...}
\end{minipage}
\vspace{-10pt}
\end{figure}}{
\begin{figure*}[t]
\begin{minipage}{.32\textwidth}
	\includegraphics[width=0.98\linewidth]{deep_interaction_ratio_bias_classes.pdf}
	\vspace{-8pt}
	\caption{Deep interaction ratios (\textit{DIRC} and \textit{DIRP}) for different bias classes and a baseline.}
	\label{fig:deep_interaction_ratio_bias_classes}
\end{minipage}
\hfill
\begin{minipage}{.32\textwidth}
    \centering
	\includegraphics[trim =12mm 0mm 4mm 0mm, width=0.98\linewidth]{cdf_of_BOTH_deep_ratio_bias_classes.pdf} 
    \vspace{-8pt}
	\caption{CDFs of \textit{DIRC} (combined) and \textit{DIRP} (public) for different bias classes.}
	\label{fig:cdf_of_both_deep_ratio_bias_classes}
    \Description{...}
\end{minipage}
\hfill
\begin{minipage}{.32\textwidth}
	\centering
	\includegraphics[trim =16mm 0mm 0mm 0mm,width=0.98\linewidth]{cdf_of_BOTH_deep_ratio_reliability_classes.pdf} 
	\vspace{-8pt}
    \caption{CDFs of \textit{DIRC} (combined) and \textit{DIRP} (public) for different reliability classes.}
    \label{fig:cdf_of_both_deep_ratio_reliability_classes}
	\Description{...}
\end{minipage}
\vspace{-10pt}
\end{figure*}
 }

 While these results may appear surprising at first, they are due to some interesting differences in the distributions.  First, note that the aggregate analysis only calculates the overall ratio 
$\frac{\sum_{i=1}^{N_c} I_{c,i}^{\text{public}}}{\sum_{i=1}^{N_c} I_{c,i}^{\text{comb}}}$ compared to the granular analysis, where the average (for example) is calculated as 
$\frac{1}{N_c} \sum_{i=1}^{N_c} \frac{I_{c,i}^{\text{public}}}{I_{c,i}^{\text{comb}}}$.  Second, taking a closer look at the distributions,  $\mathcal{D}_{Left}^{PIR}$ and $\mathcal{D}_{Right}^{PIR}$ differ substantially.  Part of this can be observed in Figure~\ref{fig:ccdf_of_combined_and_public_interactions_for_different_bias_classes}, where we can see that \textit{Left} class demonstrates notable disparities between public and combined interactions in the tail, while the \textit{Right} class exhibits relatively larger differences in the head of the distribution.  This suggests that articles with higher combined number of interactions in the \textit{Right} class may exhibit higher normalized public interaction shares, and articles with few combined interactions may exhibit relatively smaller public shares (compared to the {\em Left} class).

 To confirm this conjecture and substantiate our claim that the above differences are due to aggregated statistics being more affected by larger public interaction shares associated with the tail, we compared the public interaction ratios for articles associated with the tail and head of the distributions of the combined number of interactions. For this analysis, we split the total sets ($\mathcal{D}^{PIR}_{Left}$ and $\mathcal{D}^{PIR}_{Right}$) into two to five equally sized subsets (based on the total combined number of interactions $I_{c,i}^{comb}$ associated with each entry) and then compared the medians and averages of the public interaction ratios seen in the first and last bucket of the {\em Right} class with the corresponding values seen for the {\em Left} class.  In all cases, the \textit{Left} dominates the \textit{Right} for the head and is dominated by the {\em Right} for the tail.  For example, with three equal-sized buckets, the two classes have median ratios of 0.34 vs. 0.13 for the first bucket (head) and 0.39 vs. 0.46 for the last bucket (tail). Furthermore, we observed statistically significant differences (99\% confidence level) as per the above conjecture with both the Mann-Whitney U test (medians) and Bootstrap tests (averages), validating these differences. This confirms that the articles in the \textit{Right} class with the highest interactions have elevated normalized public interaction ratios, and those in the \textit{Left} class with the lowest interactions have similarly elevated ratios. 

 \keytakeaway{In terms of bias, the \textit{Right} class sees the smallest public interaction share among the least popular content to interact with, while the {\em Left} class sees relatively less overall public interaction share (due to its most popular articles to interact with not seeing as big public interaction share).}

\section{Deep vs. Shallow Interactions}\label{sec:deep_vs_shallow_interactions}

In this section, we investigate the prevalence of {\em deep} vs. {\em shallow} interactions and explore their variations among different reliability and bias classes.  For this analysis, 
we categorize all (emoji-based) reactions 
\CRedit{(i.e., likes, loves, sads, etc.)}{(e.g., likes, loves, sads)} 
as shallow interactions while 
 comments and shares are considered as 
deep interactions. 
This distinction, supported by the previous works \cite{aldous2019view,KIM2017441}, is motivated by the idea that comments and shares typically involve more cognitive effort and engagement from 
\CRedit{users, as they require formulating thoughts, opinions, or responses in the form of comments or actively endorsing and disseminating the content by sharing.}{users. For example, comments involve formulating thoughts and opinions, while shares actively endorse and disseminate content.} 

 To compare the depth of interaction seen for subsets of public and private posts, we 
define the following two 
 per-article metrics:\begin{itemize}
\item \textit{Deep Interactions Ratio for Combined (\textit{DIRC})}: 
 For each article, this ratio measures the proportion of the combined interactions that are classified as deep interactions.\item \textit{Deep Interactions Ratio for Public (\textit{DIRP})}: 
 For each article, this ratio measures the proportion of the public interactions that are classified as deep interactions.\end{itemize}

\CRedit
{}
{
To compute \textit{DIRC}, we divide the number of deep interactions by the total interactions for each URL. For \textit{DIRP}, we first identify the deep interactions (shares and comments) occurring on public posts associated with the URL. Summing these deep interactions across all public posts and dividing by the total number of public posts' interactions for the URL yields the \textit{DIRP}.
} 
For example, an article with 1,000 total interactions, 500 of which are deep, has a \textit{DIRC} of 0.5. If 400 interactions are public, and 100 of those are deep, the \textit{DIRP} is 0.25. We next analyze the distribution of \textit{DIRC} and \textit{DIRP} values for all articles in each class.
 
{\bf Comparisons of bias classes:}
We first compare the deep interaction patterns of the bias classes. Figure~\ref{fig:deep_interaction_ratio_bias_classes} shows the distributions of \textit{DIRC} and \textit{DIRP} for the different classes.
As a baseline, we also include the distributions for the general population (rightmost box pair), considering all samples.

First, referring to the figure, for each bias class, we observe substantial differences between the distributions of
 \textit{DIRC} and \textit{DIRP}.
\CRedit{For example, for every percentile and for the means, the DRIC distributions (combined) have significantly larger values than the DIRP distributions (public).}
{For example, for every percentile and for the means, the \textit{DIRC} distributions (combined) have significantly larger values than the \textit{DIRP} distributions (public), indicating that users engage in more substantive interactions in less-public settings. This suggests that deeper engagement—through comments and shares—tends to \textit{comparatively} have higher ratios in spaces with fewer visibility constraints, where users may feel more comfortable expressing opinions or endorsing content. This pattern is consistent across all bias (Figure~\ref{fig:cdf_of_both_deep_ratio_bias_classes}) and reliability classes (Figure~\ref{fig:cdf_of_both_deep_ratio_reliability_classes}), reinforcing the notion that private interactions are an essential component of online engagement dynamics.}
This again highlights the difference between the public and combined interactions and the importance of considering both the private and public spheres in future research. 
This observation leads us to the following key insight.
\keytakeaway{Irrespective of the news class, users tend to engage more deeply in private discussions compared to public ones.}

Now, and perhaps even more importantly, to better understand to what extent the relative level of interaction seen for different bias classes depends on which dataset was used we apply the same statistical testing methodology as discussed in previous sections for the two sets independently. First, the Kruskal-Wallis test provides evidence that the distributions differ significantly among the bias classes, regardless of using the \textit{DIRC} and \textit{DIRP}, with p-values of $7.14 \cdot 10^{-23}$ and $2.08 \cdot 10^{-9}$, respectively. 
Comparing the p-values, we note that the differences appear more significant using the combined set (\textit{DIRC}) than using only the public data (\textit{DIRP}).
Second, consistent with visual comparison in the figure, the post-hoc Dunn test only supports statistically significant differences between the \textit{Right} group (which has the biggest fraction of deep interactions regardless of using the combined or pubic sets) and the other two groups but not among the \textit{Center} and \textit{Left} group. Finally, the Mann-Whitney U test and bootstrapping support the same pattern for the median and means, with the only significant differences again being between the \textit{Right} class and the other two bias classes. For the bias classes, the main results regarding the relative depth of the interactions among the classes are, therefore, consistent regardless of the dataset.

To illustrate 
 the 
tendency of the posts with \textit{Right}-biased articles 
 seeing 
a higher deep interaction share, regardless of whether publicly shared or not, we include the CDFs of \textit{DIRC} (combined) and \textit{DIRP} (public) for all three biases classes in Figure~\ref{fig:cdf_of_both_deep_ratio_bias_classes}. We note that in both cases, the \textit{Right} class exhibits a noticeable shift to the right, supporting the following insight.

\keytakeaway{Content from the \textit{Right} bias class in both public and combined contexts gets deeper user interactions.}

{\bf Comparison of reliability classes:}
We now shift our focus to the reliability classes and their deep interaction ratios.  Figure~\ref{fig:cdf_of_both_deep_ratio_reliability_classes} summarizes these results. 
While we again observe deeper interactions for the combined set (suggesting shallower interactions with public posts), the relative differences among the reliability classes themselves are less apparent compared to those observed among the bias classes.  We also observe some smaller differences in which relative differences are significant, highlighting the value of also considering the combined data (not only the public data).

First, starting with the combined dataset (\textit{DIRC}), we observe the presence of significant differences among the distributions (Kruskal-Wallis test), with the post-hoc Dunn test (distribution) and Mann-Witney U test (median) revealing that only the \textit{Most-unreliable} class differs statistically from the others. This observation is further
 visually supported 
 in Figure~\ref{fig:cdf_of_both_deep_ratio_reliability_classes}, which clearly shows a shift to the right in the 
 \textit{DIRC} distribution of the \textit{Most-unreliable} 
 class.

Switching our focus to the \textit{DIRP} distributions, containing only public interactions, we interestingly observe a slightly different pattern, as we
here 
also observe statistical differences between some additional classes. First, as suggested by the \textit{DIRP} CDFs in Figure~\ref{fig:cdf_of_both_deep_ratio_reliability_classes},
both extreme classes (\textit{Most-unreliable} and \textit{Most-reliable}) are statistically different from the other two classes (Kruskal-Wallis test followed by Mann-Whitney test, followed by the post-hoc Dunn test). Second, no statistical significance was observed between the \textit{Most-reliable} and \textit{Most-unreliable} classes themselves or between the \textit{Reliable} and \textit{Unreliable} classes. We base the following insight based on our observations across the two datasets.
\keytakeaway{Users exhibit higher deep interaction
levels with content from the \textit{Most-unreliable} class.}

\section{\CRedit{}{Topic-Based Analysis}}\label{sec:topic_analysis}
In this section, we extend our investigation by conducting a topic-based analysis of news articles to assess whether engagement patterns vary across different content themes. 
\CRedit{To this goal,}{For this analysis,}
we first extracted the textual content from the dataset news articles, a process that succeeded for 14,996 items. We then employed the ChatGPT 4o-mini model to classify each article into one of 14 predefined topics: \emph{Politics}, \emph{Health~\&~Medicine}, \emph{Crime}, \emph{Business ~\&~Finance}, \emph{Environment~\&~ Climate}, \emph{Entertainment}, \emph{Education}, \emph{Sports}, \emph{Science}, \emph{Lifestyle \& Leisure}, \emph{Religion},  \emph{Technology}, \emph{Arts~\&~ Culture}, and \emph{Food}. This specific set of topics was selected to capture the diverse range of subject matters prevalent in contemporary news consumption, while also reflecting societal interests and trends. 

Notably, the ordering of topics here mirrors the distribution observed in our dataset—ranging from the most frequent (\emph{Politics}, with 7,956 articles) to the least frequent (\emph{Food}, with 57 articles). However, it is important to note that these frequencies are influenced by two key factors: (1) the initial article selection criteria of Ad Fontes Media, and (2) the timeframe of our dataset, which spans a period heavily impacted by the COVID-19 pandemic, leading to an elevated presence of \emph{Health \& Medicine} content (the second most prevalent topic).
Finally note that for articles where none of these predefined topics were clearly identifiable, the model was instructed to leave them uncategorized. In total, 14,650 articles (97.7\%) were successfully classified into one of the 14 categories.

First, to assess whether our earlier findings hold within specific topics, we conducted a stratified analysis across topic groups.  
\CRedit{Our results indicate that, for many cases, we see the previous patterns statistically significant.}{Our results indicate that the previously observed patterns remain, with statistical significance for several of the classes with enough samples.} 
For example, in the case of the \emph{Politics} topic the insights reported in both 
Sects.~\ref{sec:private_vs_public}
and~\ref{sec:deep_vs_shallow_interactions} remain statistically robust when considering the reliability dimension. As another example, for \emph{Health \& Medicine} (the 2nd largest topic), all the conclusions in 
Sects.~\ref{sec:private_vs_public}
and~\ref{sec:deep_vs_shallow_interactions} for both the bias and reliability dimensions are still statistically significant. These confirm the general validity of our broader findings within many of these topics. 

\begin{figure}[t]
\centering
\includegraphics[trim= 0mm 5mm 0mm 0mm, clip, width=0.98\columnwidth]{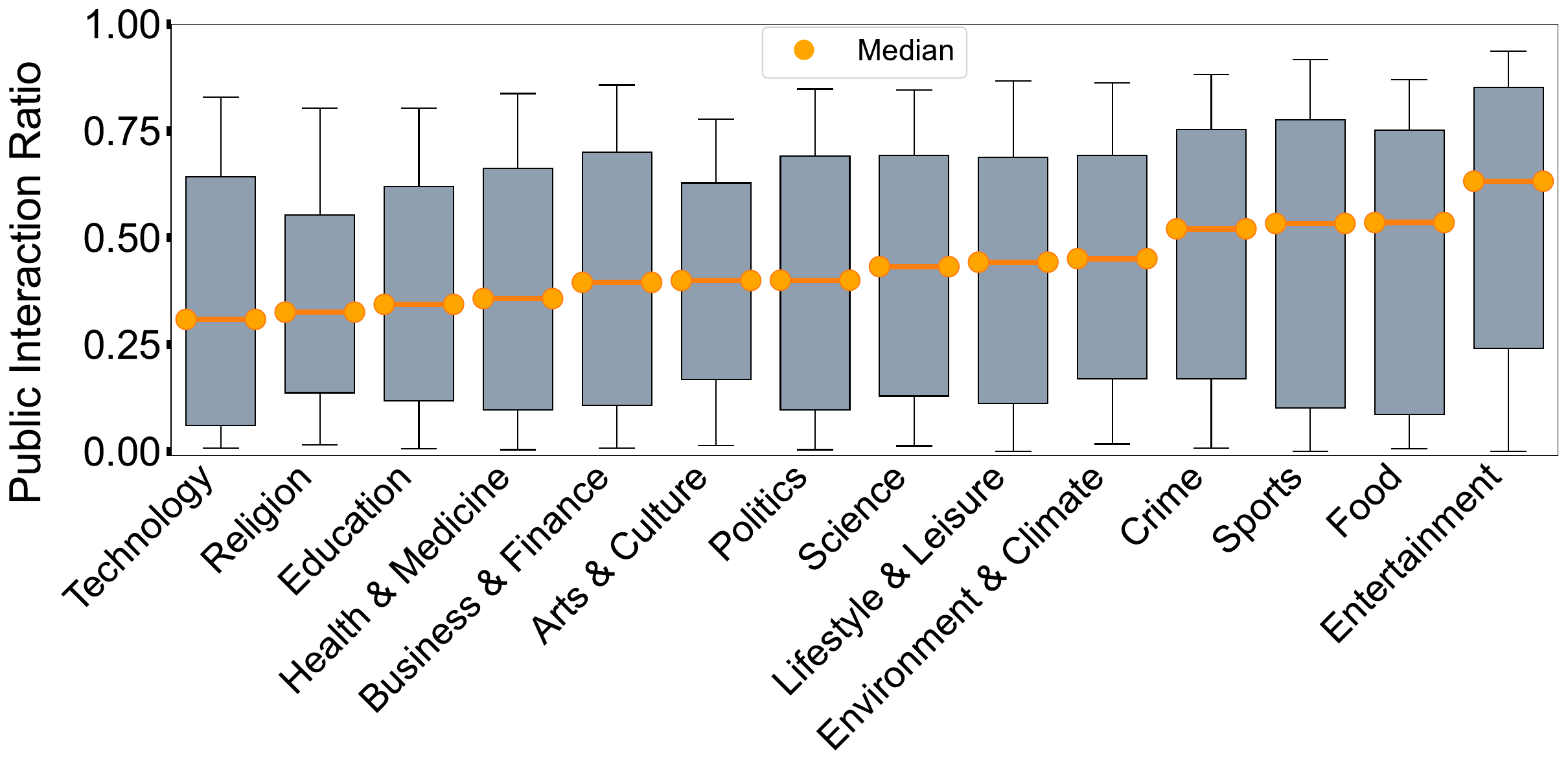} 
\vspace{-10pt}
\caption{Boxplots of the public interaction ratio for different topics (sorted by median).}
\label{fig:boxplot_of_public_interaction_share_for_each_topic}
\end{figure}

Second, to explore how engagement varies across topics, we analyzed the distributions of the public interaction ratio for each topic. This analysis was conducted independent of the articles' bias and reliability, as some topics had too few samples to allow for a breakdown by those categories. 
Figure~\ref{fig:boxplot_of_public_interaction_share_for_each_topic} shows the distribution of the public interaction ratios (as a boxplot) for each of the topics, sorted based on their medians. 
From this figure, we can first see that 
\CRedit{\textbf{\emph{Entertainment}-related articles exhibit the highest public interaction share},}{\emph{Entertainment}-related articles exhibit the highest public interaction ratio,} 
indicating that content from this category is more likely to be engaged with in public discussions. This suggests that entertainment news, which often includes celebrity updates and viral stories, is inherently more shareable and widely disseminated in public spaces.
On the other hand, 
\CRedit{\textbf{\emph{Technology}-related articles exhibit the lowest public interaction share},}{\emph{Technology}-related articles exhibit the lowest public interaction share,} 
meaning they receive a disproportionately high share of engagement in private spaces. This may be attributed to the technical nature of these articles, which are often discussed within niche communities or shared in professional or interest-based private groups.

\begin{figure}[t]
\centering
\includegraphics[trim= 0mm 5mm 0mm 8mm, clip, width=0.9\columnwidth]{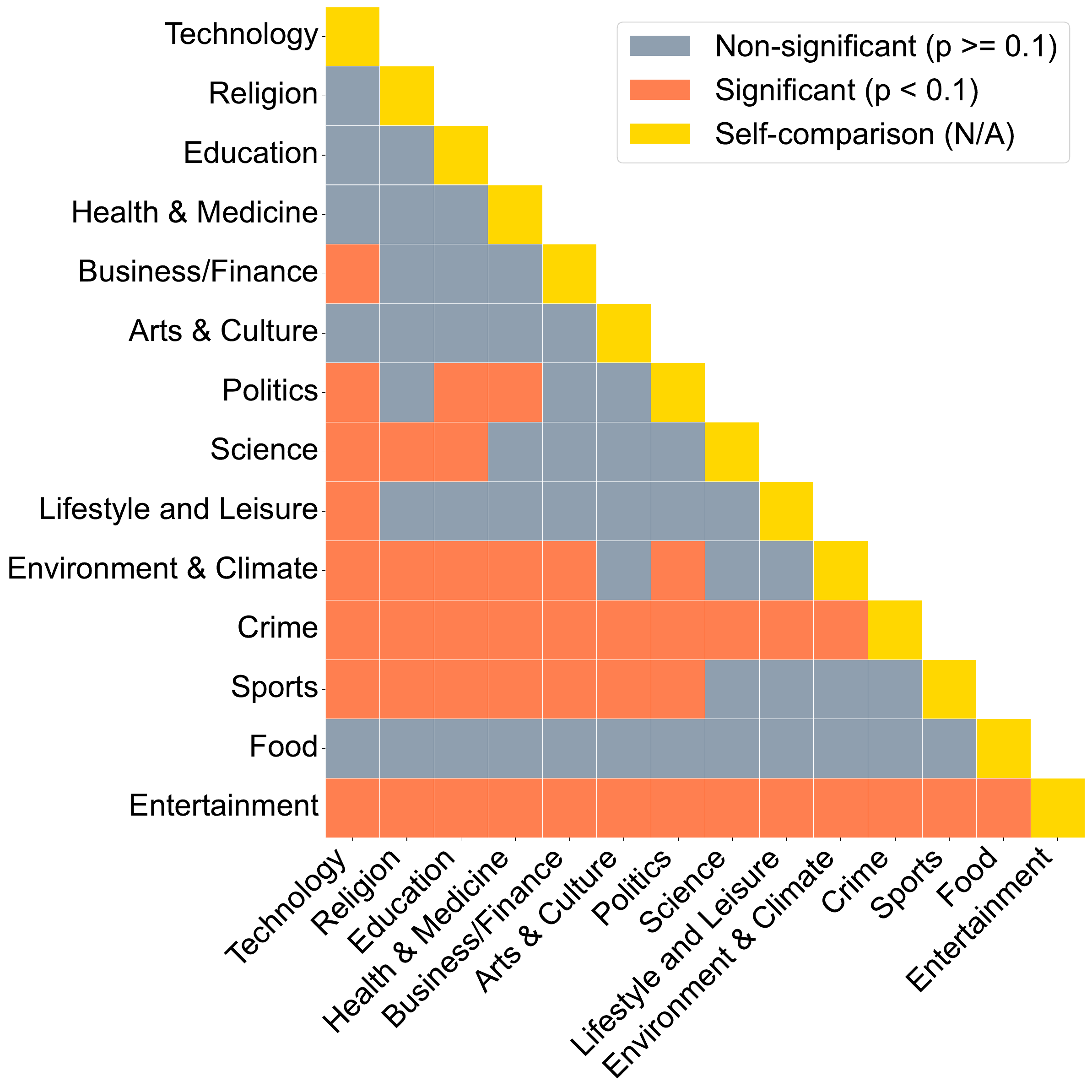}
\vspace{-8pt}
\caption{Heatmap of Kruskal-Wallis test results for topics' public interaction ratios.}
\label{fig:kruskall_wallis_heatmap}
\end{figure}

It is important to note that not all observed differences in public interaction ratios between topics in Figure~\ref{fig:boxplot_of_public_interaction_share_for_each_topic} are statistically significant. To validate the statistical significance of these observed differences, we conducted pairwise Kruskal-Wallis tests between topic groups. Figure~\ref{fig:kruskall_wallis_heatmap} presents a heatmap illustrating the statistical significance of differences in public interaction ratios across topics. Orange-colored cells here indicate significant differences ($p < 0.1$), while gray cells indicate non-significant comparisons.

This analysis confirms that \emph{Entertainment}-related content has a significantly higher public interaction share compared to all other topics, whereas \emph{Technology}-related content exhibits significantly lower public engagement compared to most categories. Conversely, topics such as \emph{Food}, we could not find statistically significant differences in public interaction ratios compared to other topics (except for \emph{Entertainment}). It is worth noting that we selected a p-value threshold of 0.1 for our analysis, as stricter thresholds (e.g., 0.01) did not yield many statistically significant differences between topics. This more lenient threshold was chosen primarily 
to compensate for the limited sample size for certain topics in our dataset.

\section{\CRedit{}{Top Publishers Analysis}\label{sec:top_publishers}}
To further contextualize our findings, we analyze the public interaction ratio (\textit{PIR}) and the deep interaction ratio (\textit{DIRC}) for the top publishers in our dataset. In this context, ``top publishers'' refers to those with the highest number of articles in our (and Ad Fontes Media) dataset. For this analysis, we excluded Yahoo, as it primarily functions as a news aggregator and does not produce original content.

\begin{figure*}[t]
    \centering
    \begin{subfigure}{0.45\textwidth}
        \centering
        \includegraphics[trim= 0mm 5mm 0mm 0mm, clip, width=\textwidth]{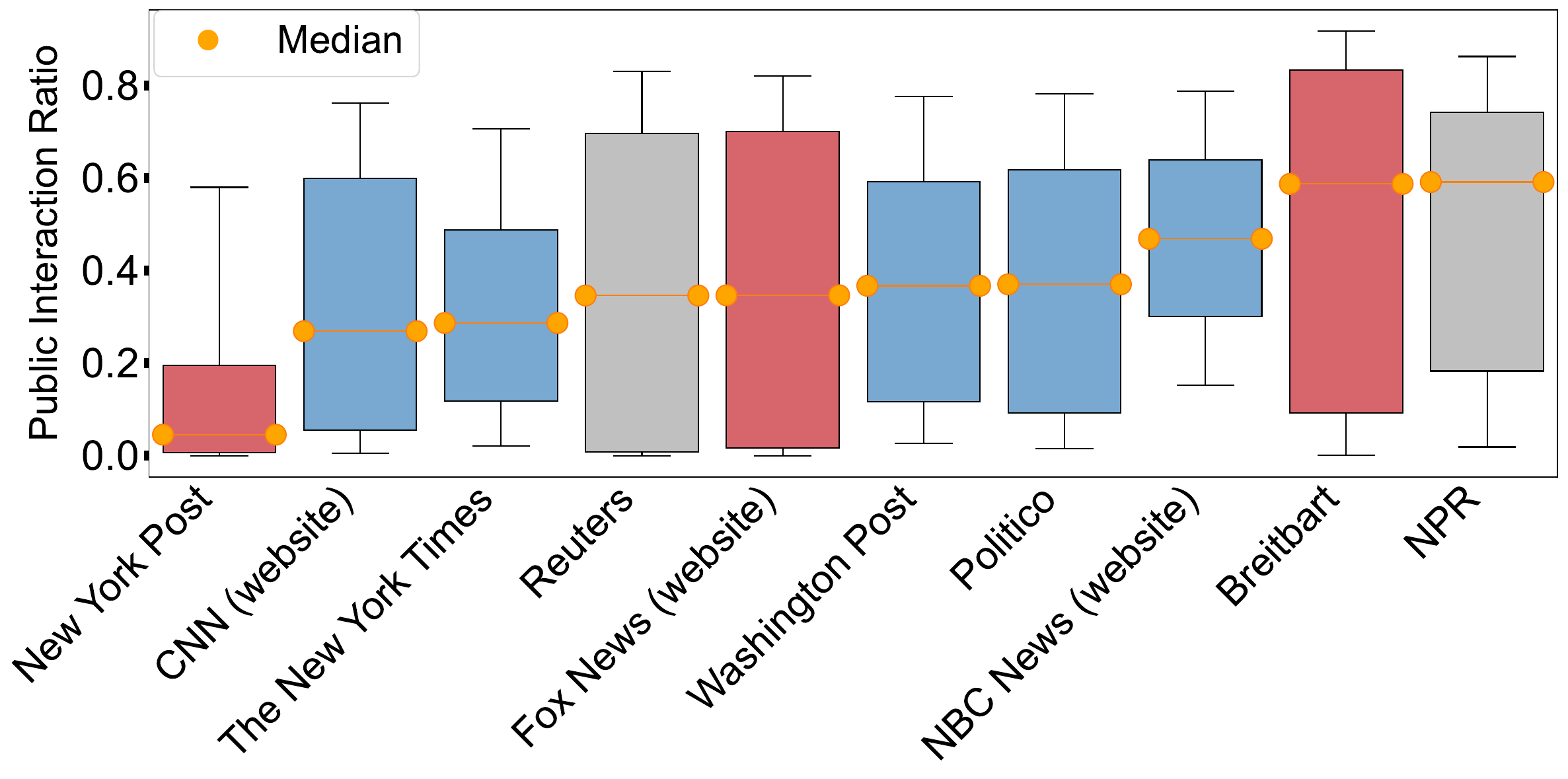}
        \caption{Boxplots of the public interaction ratios (sorted by median).}
        \label{fig:boxplot_of_public_interaction_share_for_top_10_monikers}
    \end{subfigure}
    \hfill
    \begin{subfigure}{0.45\textwidth}
        \centering
        \includegraphics[trim= 0mm 5mm 0mm 0mm, clip, width=\textwidth]{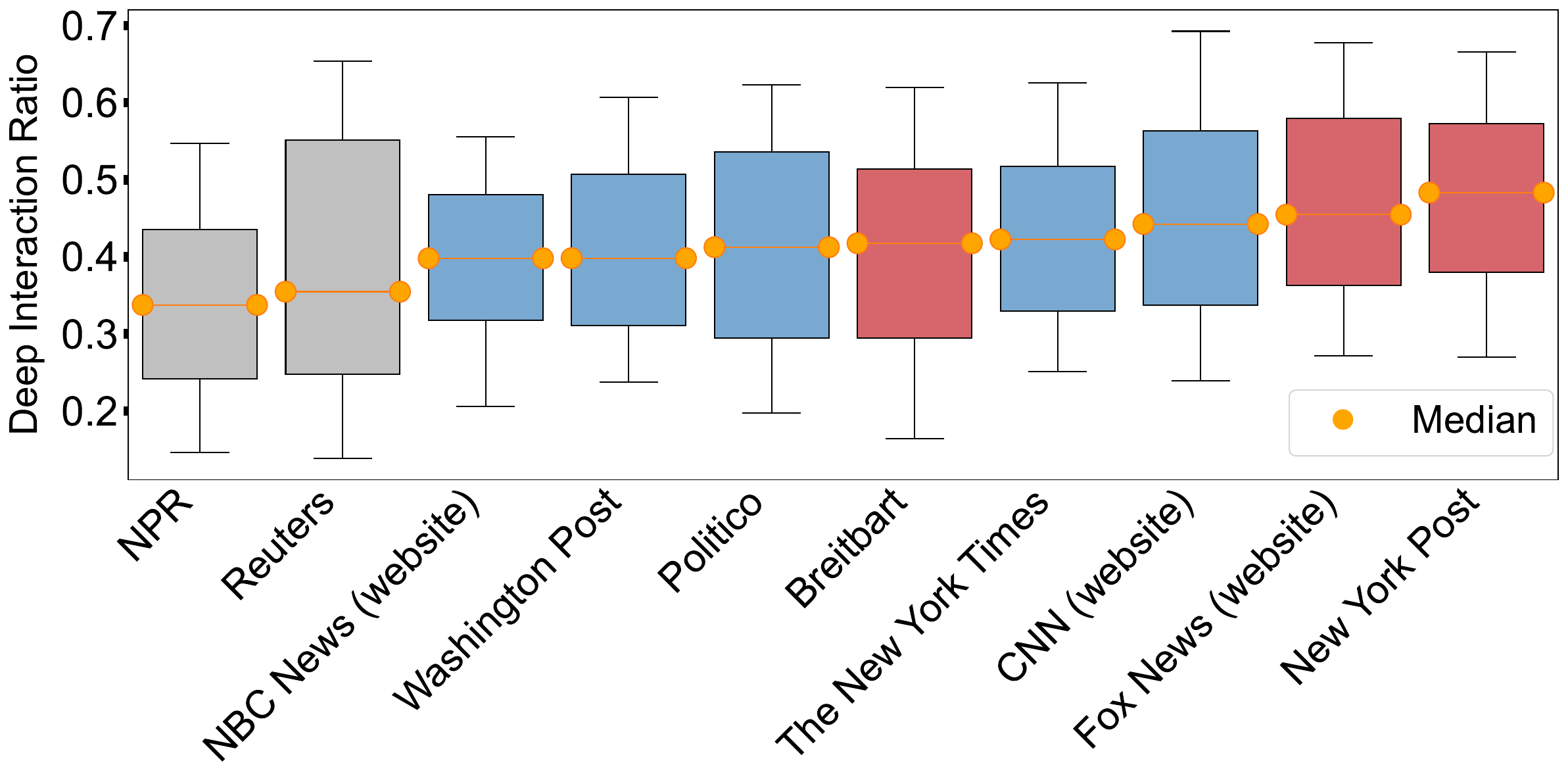}
        \caption{Boxplots of the deep interaction ratios (sorted by median).}
        \label{fig:boxplot_of_deep_interaction_ratio_for_top_10_monikers}
    \end{subfigure}
    \caption{Comparison of (a) public interaction ratio and (b) deep interaction ratio for the top-10 outlets.}
    \label{fig:combined_boxplots}
\end{figure*}

We begin by examining the public interaction ratio across the top publishers. Figure~\ref{fig:boxplot_of_public_interaction_share_for_top_10_monikers} presents a boxplot of the public interaction ratio for each outlet, sorted by median values. For easier interpretation, the color of each ``box" indicates the bias class (i.e., \textit{Left}, \textit{Center}, or \textit{Right}) associated with each publisher (based on the labeling of the articles they have published). It is worth noting that some outlets, such as \textit{Politico} and \textit{NBC News}, may shift between the \textit{Left} and \textit{Center} classifications over time due to editorial changes or evolving media landscapes. From this figure, we observe that \textit{New York Post} exhibits the lowest public interaction ratio, suggesting that a larger proportion of its engagement occurs in private spaces. This could be attributed to the nature of its audience, which may prefer engaging with its content in more private settings, such as closed groups. Additionally, the sensationalist reporting of the \textit{New York Post} might encourage discussions that users feel more comfortable having in less-visible online spaces.

Next, we investigate which of these publishers generate the deepest user engagement, as measured by the deep interaction ratio (\textit{DIRC}). Figure~\ref{fig:boxplot_of_deep_interaction_ratio_for_top_10_monikers} illustrates the \textit{DIRC} distributions for the top publishers, again sorted by median values. Notably, \textit{New York Post} and \textit{Fox News}, both from the \textit{Right} class, exhibit the highest median \textit{DIRC} values, followed by \textit{CNN} and \textit{New York Times} from the \textit{Left} class. At the lower end of the spectrum, we find \textit{NPR} and \textit{Reuters}, both from the \textit{Center} class. In general, we observe that publishers with the highest \textit{DIRC} values tend to be biased ones, while those from the \textit{Center} class have the lowest values. The probability of this pattern occurring if we randomly sort the 10 publishers is approximately 2.22\%. This observation may suggest that users are more likely to engage deeply with content from publishers that have a clear political leaning, compared to those that maintain a more neutral stance.

.

\section{Related Works}\label{sec:related_works}

 This research most closely relates to studies exploring user engagement dynamics with various social media content. Our main contribution lies in the novel analysis of interaction disparities between public and private news article sharing on Facebook, particularly concerning articles with varying bias and reliability. 
Examining interactions depth adds another layer to our analysis and findings.

There exist studies 
\CRedit{that partially touch upon some of the dimensions highlighted in this paper.}{that partially address some dimensions highlighted in this paper.} 
However, as discussed in the introduction, the vast majority of these works, particularly pertaining to Facebook, predominantly concentrate on the platform's public sphere. 
To contextualize our study within the context of the existing literature and to highlight 
the novelty of our contribution, we next describe the most closely related works.

One line of this prior research has studied the differences in the engagement levels of different bias or reliability (or both) classes on different platforms including Facebook~\cite{edelson2021understanding,Understanding_Audience_Engagement_with_Mainstream,people_turned_to}, Reddit~\cite{weld2021political,10.1093/pnasnexus/pgad018}, and 
\CRedit{Twitter\cite{Vosoughi2018TheOnline}}{Twitter\cite{mohammadinodooshan2024understanding,Vosoughi2018TheOnline}}. For example, Edelson et al.~\cite{edelson2021understanding}, 
while focusing on the public sphere, 
conducted a large-scale study on user engagement with 7.5 million posts from 2,551 publishers (Facebook pages) across both bias and reliability dimensions. Their results show that individual posts from non-misinformation news outlets tend to attract lower median engagement than misinformation,  aligning with our findings. Weld et al.~\cite{weld2021political} carried out a similar investigation on Reddit rather than on Facebook. 
When considering the reliability parameter, Vosoughi et al.~\cite{Vosoughi2018TheOnline} analyzed a dataset of approximately 126,000 stories tweeted by over 3 million people more than 4.5 million times, finding that false news stories reached more people than true news and diffused significantly farther, faster, deeper, and more broadly than the truth across various categories of information. Notably, during the critical period of the COVID-19 pandemic and by doing a cross-national study, Altay et al.~\cite{people_turned_to} delineated the surge in online news consumption, with credible news outlets witnessing a significant boost. Samory et al.~\cite{samory2020characterizing} characterized the social media news sphere through user co-sharing practices by focusing on 639 news sources, both credible and questionable, and divided them into 4 clusters, and characterizing them according to the audience that shares their articles on Twitter and how the stylometric features used by each cluster is successful in getting users engagement.
Lamot et al.~\cite{lamot2022we} examined how news headlines are remediated on Facebook and how this affects user engagement. Finally, Boukes et al.~\cite{boukes2022comparing} compared user-content interactivity and audience diversity across news and satire, finding differences in online engagement between satire, regular news, and partisan news. Considering the bias dimension, on the other hand, Wischnewski et al.~\cite{Wischnewski2021ShareworthinessTwitter} discovered that users are more inclined to share hyperpartisan news 
articles that coincide with their own political views. The bias towards the right party in sharing and engagement with news is another thesis that has been studied by previous works on both Twitter~\cite{10.1093/pnasnexus/pgac137} and Facebook~\cite{gonzalez2023asymmetric}.

Research has also examined other dimensions of user engagement, with, for example, Aldous et al.~\cite{aldous2019view} showing that content topics influence engagement. Studies have found that emotional content generates higher Facebook engagement, with Maier et al.~\cite{compassion_fatigue} demonstrating this effect while controlling for author influence. Finally, some works have considered the time factor and analyzed temporal patterns of engagement with different news classes~\cite{vassio2022mining}.

In summary, while previous research has partly explored some dimensions of user engagement with news, our study is the first to directly compare the dynamics of public and private sharing of news articles on Facebook.
\section{Conclusions}
\label{sec:conclusions}

In conclusion, this study makes several important contributions to 
the
understanding of news engagement on social media. By developing a robust methodology through collaboration with the CrowdTangle team, we present the first comprehensive comparison of engagement dynamics between Facebook's publicly tracked content and overall platform interactions. Our analysis of over 19K news articles reveals that public engagement patterns often fail to reflect platform-wide behavior, with significant implications for research methodology and content moderation strategies.

Our findings challenge several assumptions about news engagement on Facebook. First, users consistently engage more deeply with content in less public spaces, regardless of the content's bias or reliability classification. Second, particular attention should be paid to right-biased and least reliable news content, which generate notably deeper engagement across both public and less public spheres. Third, we highlight some important limitations of focusing solely on Facebook's public sphere. For example, the extremities of the reliability spectrum ({\em Most-reliable} and {\em Most-unreliable}) exhibited a tendency to elicit deeper interactions from users within the public sphere. However, this narrative shifted when examining combined interactions, indicating that only the {\em Most-unreliable} class triggered statistically significant deeper engagement from Facebook users.  This finding suggests that research conclusions drawn solely from public data may not accurately represent overall platform engagement patterns, as not only the magnitude of engagement differs between public and private but also the relative ordering of how different content types engage users.

These insights carry significant implications for multiple stakeholders. For researchers, our findings underscore the importance of considering engagement disparities between public and less public spheres when studying social media behavior, particularly for news content at different ends of the bias and reliability spectrums. For content moderators, our results suggest the need to reevaluate strategies based solely on publicly visible interactions. For policymakers, our findings highlight the importance of considering both public and less public spheres when developing policies for social media governance.

Future research should investigate the underlying factors driving these engagement patterns, particularly through user interviews,
\CRedit{and behavioral studies.}{behavioral studies, and analysis of post comments.} 
While challenges remain in understanding social media dynamics, our findings illuminate significant differences in how users interact with news across Facebook's diverse spheres, highlighting the complexity of news-sharing behavior on social media platforms.

\begin{acks}
The authors express their gratitude to CrowdTangle for providing the Facebook data. They also extend their thanks to the anonymous reviewers for their insightful comments that helped improve the paper.
This work was partially supported by the Wallenberg AI, Autonomous Systems and Software Program (WASP) funded by the Knut and Alice Wallenberg Foundation.
\end{acks}

\bibliographystyle{ACM-Reference-Format}
\bibliography{references}
\end{document}